%% file: gubbiotti_rev_eof.tex
\documentclass[]{interact}

\usepackage{color}
\usepackage{epstopdf} 
\usepackage[caption=false]{subfig} 

\usepackage[numbers,sort&compress]{natbib}
\bibpunct[, ]{[}{]}{,}{n}{,}{,}
\makeatletter
\def\NAT@def@citea{\def\@citea{\NAT@separator}}
\makeatother

\theoremstyle{plain}

\theoremstyle{definition}

\theoremstyle{remark}

\usepackage{lineno}

\begin{document}

\title{Electroosmosis in nanopores: Computational methods and technological applications}

\author{
\name{Alberto Gubbiotti\textsuperscript{a}
\thanks{CONTACT Alberto Gubbiotti. E-mail: alberto.gubbiotti@uniroma1.it 
Mauro Chinappi. E-mail: mauro.chinappi@uniroma2.it}, Matteo Baldelli\textsuperscript{b}, 
Giovanni Di Muccio\textsuperscript{b}, 
Paolo Malgaretti\textsuperscript{c}, 
Sophie Marbach\textsuperscript{d,e}, 
Mauro Chinappi\textsuperscript{b}}
\affil{
\textsuperscript{a}Dipartimento di Ingegneria Meccanica e Aerospaziale, Sapienza Universit\`a di Roma, 00184 Roma, Italia; 
\textsuperscript{b}Dipartimento di Ingegneria Industriale, Universit\`a  di Roma Tor Vergata, 00133 Roma, Italia; 
\textsuperscript{c}Helmholtz Institute Erlangen-N{\"u}rnberg for Renewable Energy (IEK-11), Forschungszentrum J{\"u}lich, Cauer Stra{\ss}e 1, 91058 Erlangen, Germany;
\textsuperscript{d}Courant Institute for Mathematical Sciences, New York University, New York, 10012, USA;
\textsuperscript{e}CNRS, Sorbonne Universit{\'e}, Physicochimie des Electrolytes et Nanosystèmes Interfaciaux, F-75005 Paris, France; 
} 
}

\maketitle

\begin{abstract}
Electroosmosis is a fascinating effect where liquid motion is induced by an applied electric field.  Counter ions accumulate in the vicinity of charged surfaces, triggering a coupling between liquid mass transport and external electric field. In nanofluidic technologies, where surfaces play an exacerbated role, electroosmosis is thus of primary importance.  Its consequences on transport properties in biological and synthetic nanopores are subtle and intricate. Thorough understanding is therefore challenging yet crucial to fully assess the mechanisms at play. Here, we review recent progress on computational techniques for the analysis of electroosmosis and discuss technological applications, in particular for nanopore sensing devices.
\end{abstract}


\input{introduction.tex}

\input{continuum.tex}
\input{atomistic.tex}

\input{mesoscale.tex}
\input{techno.tex}

\section*{Acknowledgements}
This research is part of a project that has received funding from the European Research Council (ERC) under the European Union’s Horizon 2020 research and innovation programme (grant agreement No. 803213).
The authors are grateful for fruitful discussions with 
Aleksandar Donev, Brennan Sprinkle and Francesco Viola. 
The authors would also like to thank the researchers
 that provided us high quality figures,
S.L. Zhang Y. Yao for Fig.~\ref{fig:mechanism}h,
G. Maglia and K. Willems for Fig.~\ref{fig:cont}a,d,
M. Gracheva for Fig.~\ref{fig:cont}b-c,
A. Aksimentiev for Fig.~\ref{fig:biological}c-f.
S.M. acknowledges funding from the MolecularControl project, 
European Union’s Horizon 2020 research and innovation programme 
under the Marie Sk\l{}odowska-Curie grant award number 839225. 
S.M. was supported in part by the MRSEC Program of the 
National Science Foundation under Award Number DMR-1420073.
MB was supported by Università Italo Francese under Vinci grant (project C3-1352).
PM acknowledges funding by the Deutsche Forschungsgemeinschaft (DFG, 
German Research Foundation) Project-ID 416229255 SFB 1411.
GDM and MC acknowledges the computational resources provided by the  
CSCS Swiss National Supercomputing Centre (Project-ID s1026) and 
Italian CINECA (Projects IsB21\_FLOWYAX and IsC86\_CSGGEOF)Italian.

\bibliography{biblio_merge}
\bibliographystyle{unsrt}

\end{document}

%% file: introduction.tex
\vspace{0.2 cm}

\section{Introduction}

In the early 19th century, 
two independent experiments by 
Ferdinand Friedrich Reuss and Robert Porrett Jr. 
identified a curious yet remarkable phenomenon: when an electric current
flows between two compartments containing an electrolyte solution separated by a porous membrane,
a net flow of solution
builds up~\cite{reuss1809notice,porrett1816curious,biscombe2017discovery}.
Today, in the literature, 
the net transport of an electrolyte solution induced by an external electric field
is commonly referred to as \textit{electroosmosis}.
Electroosmosis is often used to actuate fluids in micro and nano 
fluidic devices~\cite{micronanofluid,haywood2015fundamental} and plays a major role in determining the ionic 
conduction properties of nanoscale systems~\cite{yusko2010electroosmotic,balme2015ionic}.
It is especially key in the context of nanopore sensing 
technologies~\cite{chinappi2018protein,bayley2000stochastic,xue2020solid}
where a voltage is applied between two 
reservoirs communicating via a single nanopore and the measured current is used to infer 
properties of the analytes translocating through (or interacting with) the nanopore. 
In this context, electroosmosis holds great promise for controlling analyte
capture~\cite{chinappi2020analytical,saharia2021modulation,boukhet2016probing,huang2017electro,asandei2016electroosmotic}
and translocation~\cite{ermann2018promoting,hsu2016manipulation}.

\subsection{Electroosmosis working principle}
\label{sec:eofprinciple}

Micro and nanofluidic systems are intrinsically inhomogeneous 
due to the presence of confining walls, and, in many applications, 
the fluid is a liquid solution containing neutral and charged chemical species. 
The inhomogeneities due to wall geometry, chemical composition and charge 
affect the concentration of all the dissolved species, 
potentially inducing local charge accumulation even in a fluid which 
is globally neutral. 
An electroosmotic flow (EOF) arises when an external electric field 
acts on 
such an inhomogeneous system, 
\textit{e.g.} when a voltage drop $\Delta V$ is applied
at the two ends of a pore.
The electric field exerts a net force 
on the charged portions of fluid
that, in turn, sets the fluid in motion. 

\begin{figure}[t]
\includegraphics[width=\linewidth]{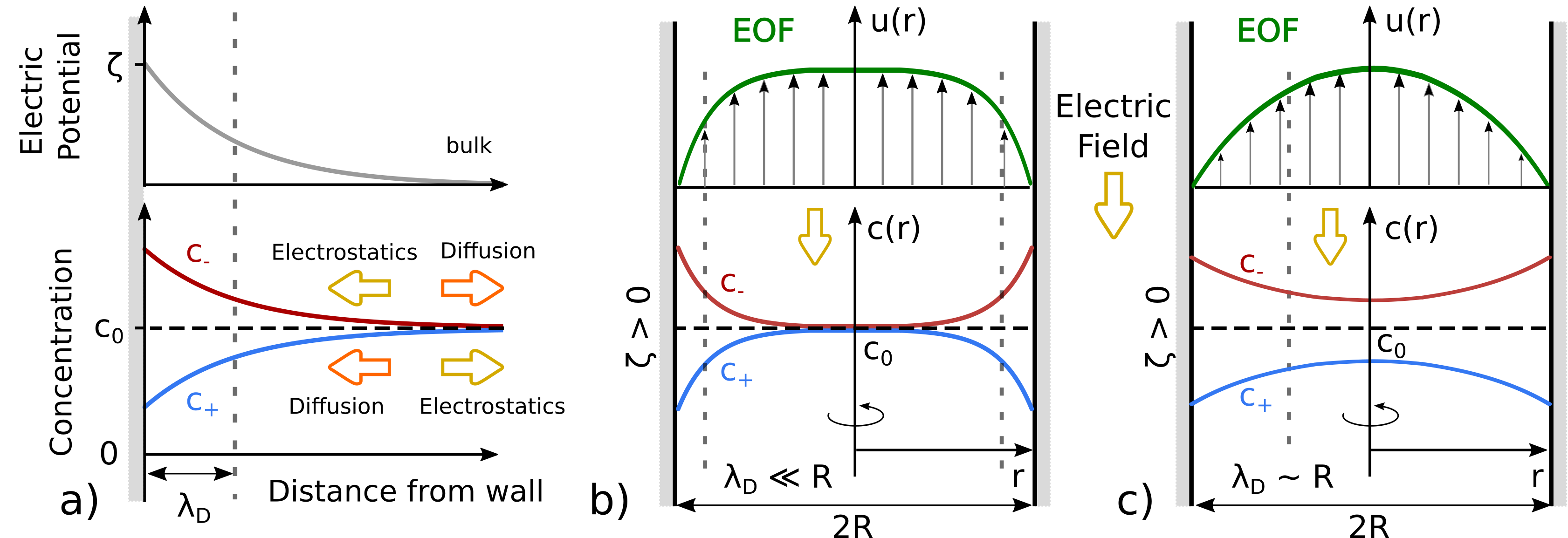}
\caption{
	{\bf Electroosmosis working principle.}
	{\bf a)} Ionic density profiles $c_+$ and $c_-$ close to a 
	planar wall with given electric wall potential, $\zeta$.
	The $\zeta$-potential induces an accumulation or depletion 
	of ions according to their charge 
	(here positive ions are in blue and negative ions in red). 
	The accumulation is larger near the wall, 
	and decreases far from the wall 
	with a characteristic length $\lambda_D$, 
	known as the Debye length. 
	As a consequence of the net charge close to the wall, 
	the fluid may be set in motion by an external electric field parallel 
	to the wall,
	as shown in {\bf b} and {\bf c}.
	{\bf b)} Electroosmotic flow (EOF) in a nanopore (no-slip boundary condition) for
	non-overlapping Debye layer, pore radius $R$ large compared to 
	the Debye length $\lambda_D$. 
	{\bf c)} EOF 
	for overlapping Debye layers, $R \simeq \lambda_D$.
}
\label{fig:sketch}
\end{figure}

A pictorial view of the simplest possible system 
in which electroosmosis occurs is sketched in 
Fig.~\ref{fig:sketch}a. 
Consider an ionic solution in contact with 
a planar wall. The electric potential of the wall's surface, with respect to the bulk fluid potential, termed henceforth the $\zeta$-potential, induces inhomogeneity in the system.
For simplicity, 
we assume here that the solution is globally neutral far from the wall, 
with only two ionic species (anions and cations) 
with equal valency and bulk concentration $c_0$. 
This is a quite common situation, easily accessible experimentally, 
for instance by dissolving a
salt such as KCl in water.
Due to electrostatic interactions with the wall, 
local electroneutrality will be broken in the near wall region
and ions will be repelled by or attracted to the wall 
depending on their charge.
Ionic diffusion balances this repulsion/attraction and tends to homogenize  
ion concentrations, see Fig~\ref{fig:sketch}a. 
This competition results in ionic accumulation/depletion 
peaked over a thin layer near the wall, referred to here as Debye 
layer\footnote{
	In the literature on charge accumulation
	at liquid-solid interfaces, 
	 different terms are used 
	to describe specific phenomena occurring 
	in this thin interfacial layer. 
	Examples are the Stern layer, 
	the Gouy-Chapman diffuse layer and 
	the electric double layer.
	In this review, 
	we do not need to enter in the specific definitions of these layers 
	and, 
	in line with part of the recent literature and
	textbooks~\cite{theoretical_microfluidics},
	we refer to the region in which the 
	ions accumulate/deplete as the Debye layer.
	}.

The balance between electrostatic interactions and diffusion 
has been quantified independently by Gouy and Chapman, 
at the onset of the 20th century~\cite{gouy1910,chapman1913}.
In their framework, it is assumed that ion concentrations verify the Boltzmann distribution 
in which the only force acting on the ion is due to the local electric field. 
In combination with Poisson's equation for electrostatics, 
a closed equation for the electric potential is obtained~\cite{fogolari2002,micronanofluid,theoretical_microfluidics}.
The Poisson-Boltzmann equation can be analytically solved 
for simple geometries, such as planar walls or channels, in its linearized form, 
which holds for sufficiently small $\zeta$-potentials, 
-- as shown by Debye and H\"{u}ckel in 
1923~\cite{debye1923,theoretical_microfluidics}.
Within this limit, the decay of the electrostatic potential in the fluid is exponential, with a 
lengthscale equal to the Debye length, here given by
\begin{equation}
\lambda_D
=
\sqrt{
	\frac{\varepsilon_0 \varepsilon_r k_BT}{q^2\sum\limits_\alpha c_\alpha Z_\alpha^2} 
}
\; ,
\label{eq:lambdaD}
\end{equation}
where $T$ is temperature, $k_B$ the Boltzmann constant, 
$\varepsilon_0$ vacuum permittivity, 
$\varepsilon_r$ relative permittivity of the liquid, 
$q$ the elementary charge and $c_\alpha$ and $Z_\alpha$ 
are respectively the number concentration (particles/volume) and valency of the 
ionic species $\alpha$ in solution, \textit{e.g.} $Z_{\alpha} = +1$ for $K^+$ and $Z_{\alpha} = -1$ for $Cl^-$.
The balance between electrostatic 
and diffusive effects is clear in Eq.~\eqref{eq:lambdaD}.
A broader $\lambda_D$ is obtained at higher temperatures.
In contrast, sharper ionic distributions are obtained when 
electrostatic interactions are enhanced, \textit{e.g} with higher valency and ion concentration, or lower relative permittivity.
In typical nanopore applications~\cite{smeets2006salt,ma2019nanopore,huang2017electro,willems2020accurate,betermier2020single}, 
with KCl water solutions at $300$~K, 
$\lambda_D$ calculated from Eq.~\eqref{eq:lambdaD} ranges from 
$~0.3\,\mathrm{nm}$ for $1\,M$ KCl 
to $10\,\mathrm{nm}$ for $1\,mM$ KCl.

The relative scale of the Debye layer $\lambda_D$ compared with the radius $R$ of the pore controls the characteristics of the EOF. When $\lambda_D \ll R$, as in most micrometric systems, see Fig.~\ref{fig:sketch}b,
the ionic distribution and the electric potential 
reach their bulk values over most of the pore volume
but in the Debye layers.
This condition is usually referred to as 
\textit{non-overlapping} Debye layers. 
When an electric field parallel to the channel axis is applied (yellow arrow), 
the net force acts only on the thin charged layers near the wall, 
generating a plug-like velocity profile,
see $u(r)$ in Fig.~\ref{fig:sketch}b. 
In such microfluidic settings, with non-overlapping Debye layers, electroosmosis enables in particular electroosmotic pumping~\cite{wang2009electroosmotic}.
In nanopores, in contrast, the channel size is often comparable to the
Debye length,
$\lambda_D \sim R$, see Fig.~\ref{fig:sketch}c.
In this case, 
the ion concentrations do not reach $c_0$ 
in the center of the channel and a net charge is present 
in the entire pore volume (\textit{overlapping} Debye layers).
Accordingly, for overlapping Debye layers $c_0$ has to be interpreted as 
the concentration of the salt in a large reservoir 
in equilibrium with the confined system.
Consequently, 
an external electric field will result in a volume force 
through the entire pore generating a velocity profile 
qualitatively similar to the parabolic Poiseuille profile
due to a pressure gradient,
see $u(r)$ in Fig.~\ref{fig:sketch}c.
For strongly overlapping Debye layers, 
recent theoretical results 
suggest that fluid charges may not be sufficient to balance surface charges, resulting in a breakdown of 
the electroneutrality condition~\cite{green2021conditions,noh2020ion}.
Overlapping Debye layers are quite typical in biological nanopores where
the pore diameter is of the order of a few
nanometers~\cite{bayley2000stochastic,willems2020accurate,asandei2016electroosmotic,boukhet2016probing}, 
while solid state nanopores,
whose size may range from subnamometer 
scale~\cite{dong2017discriminating}
to decades of nanometers~\cite{ying2018formation,won2008shrinkage,cressiot2012protein}, 
may fall in both overlapping and non-overlapping cases.

\subsection{Three routes to charge accumulation at pore walls}
\label{sec:3routes}
\begin{figure}
	\centering
	\includegraphics[width=0.9\linewidth]{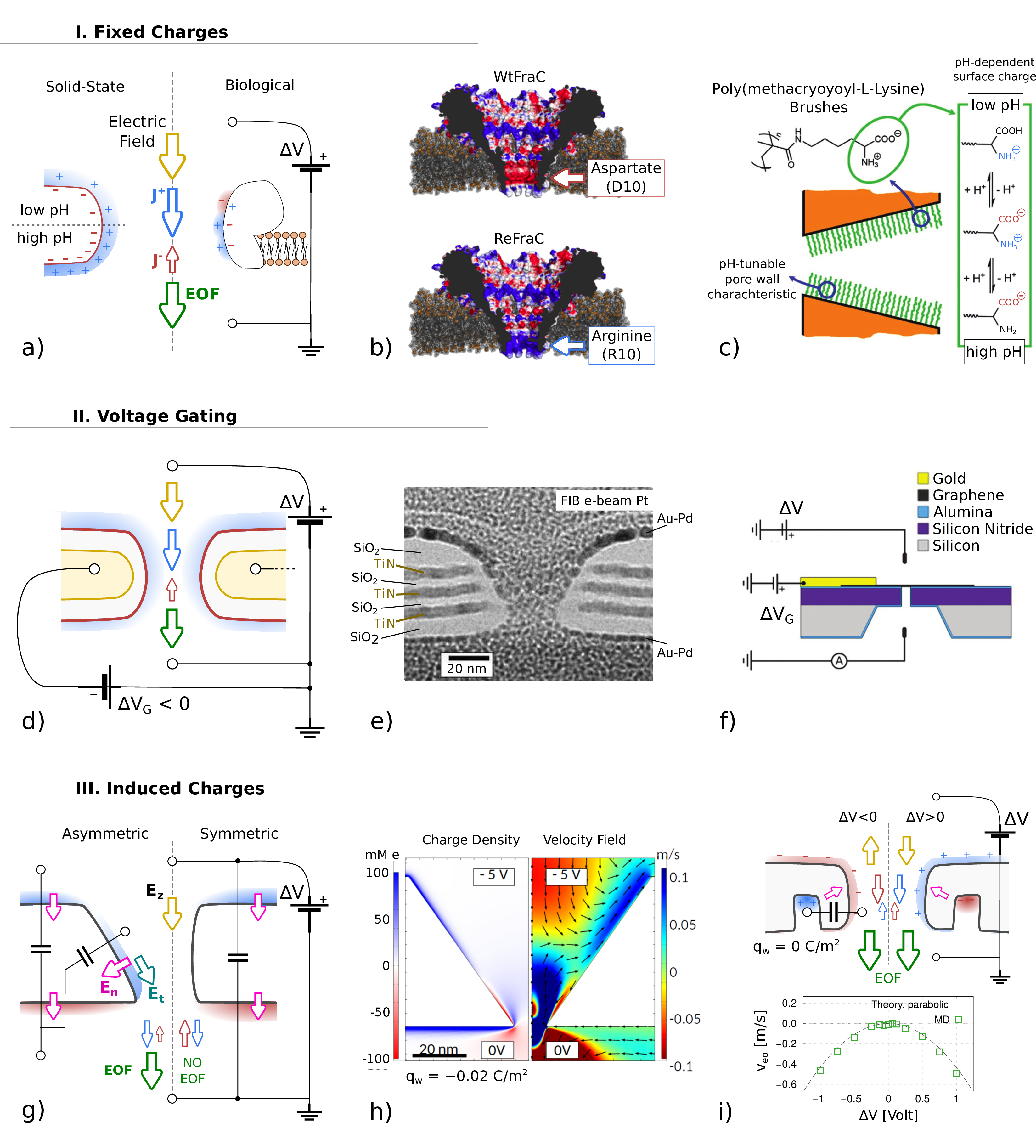}
	\caption{
		{\bf Three routes to charge accumulation at pore walls.}
		{\bf I. Fixed charges.}
		{\bf a)} Typical 
		nanopore setup.
		Surface charges at the pore wall
		attract a cloud of counterions.
		A voltage drop $\Delta V$ 
		applied across the membrane containing the pore
		generates an electric current, mainly due to the counterions, and sets the fluid in motion. 
		{\bf b)} Surface charges of a wild type
		(WtFraC) and engineered (ReFraC) transmembrane FraC protein, 
		at neutral pH. In red and blue, acidic (negative) and basic (positive) surface residues.
		The negative constriction of the
		wild type (WtFraC, cation selective) 
		is inverted into a positive one 
		(ReFraC, anion selective) 
		by a point mutation~\cite{huang2017electro}. 
		{\bf c)} Polymer brush functionalized nanochannel. 
		The charge of the coating polymer	
		can be tuned by the solution pH~\cite{yameen2009single}.
		{\bf II. Voltage gating.} 
		{\bf d)} Polarization of the membrane surface 
		is controlled by an embedded electrode,
		whose potential $\Delta V_G$ is externally 
		controlled. In this example $\Delta V_G < 0$
		and $\Delta V > 0$ for cation selectivity.
		{\bf e)} Cross-sectional Transmission Electron Microscopy 
		image of a solid state nanopore 
		showing well-separated metal levels (TiN) approaching the edge 
		of the nanopore~\cite{bai2014fabrication}. 
		{\bf f)} Single-layer suspended graphene nanopore. 
		A gating gold electrode has been patterned on graphene to control
		the electric potential of the graphene sheet~\cite{cantley2019voltage}.
		{\bf III. Induced Charge.}
		{\bf g)} Charge accumulation
		and fluxes for two different geometries.
		The normal component of the electric field
		at the solid-liquid interface (pink $\mathbf{E_n}$),
		generated by an external transmembrane voltage drop
		$\mathrm{\Delta V}$, drives the formation of the Induced Debye Layer.
		The tangential component of the electric field at the wall, 
		$\mathbf{E_t}$, 
		moves the accumulated charges.
		{\bf h)} Continuum simulation results for a negatively charged
		truncated-conical nanopore~\cite{yao2020induced}, 
		showing the distribution of the net charge concentration
		and corresponding electroosmotic velocity field at negative
		bias $\Delta V=-5~\mathrm{V}$.
		{\bf i)} Induced charge selectivity and EOF 
		in an uncharged cylindrical nanopore~\cite{di2021geometrically}, 
		exploiting symmetry breaking due to a lateral cavity surrounding 
		the nanopore. The simultaneous inversion 
		of ionic selectivity and electric field direction 
		causes a unidirectional parabolic EOF.
		Points refer to molecular dynamics simulations, the line to 
		a theoretical prediction based on continuum arguments.
		Images adapted from:
		{\bf b.} Huang et al.~\cite{huang2017electro},
		{\bf c.} Yameen et al.~\cite{yameen2009single},
		{\bf e.} Bai et al.~\cite{bai2014fabrication},
		{\bf f.} Cantley et al.~\cite{cantley2019voltage},
		{\bf h.} Yao et al.~\cite{yao2020induced},
		{\bf i.} Di Muccio et al.~\cite{di2021geometrically}.
}
	\label{fig:mechanism}
\end{figure} 

As a key ingredient for electroosmosis, 
different mechanisms have been exploited
to control the electric potential at the pore walls and hence charge accumulation.
Here, we present three methods commonly used in nanopore technology: fixed charges,
voltage gating and induced charges.  

\paragraph*{Fixed charges.} The most commonly used approach 
is based on the manipulation of fixed surface charges
at the pore wall, see Fig.~\ref{fig:mechanism}.I.
Fixed surface charges attract counterions in solution
resulting in an intrinsic accumulation of
net charge in the pore lumen.
Fixed charges are naturally present 
both in biological and solid state nanopores,
Fig.~\ref{fig:mechanism}a.

In biological pores, 
surface charges are due to acidic and basic 
titratable amino acids 
(containing carboxyl or amino functional groups) and
they often result in complex surface charge patterns 
that may include both positive and negative patches,
Fig.~\ref{fig:mechanism}b.
The charge of these amino acids 
can be partially tuned by altering 
the pH to modify their protonation
state~\cite{asandei2016electroosmotic,bonome2017electroosmotic,gu2001prolonged}.
Moreover, biological pores can be engineered with 
point mutations~\cite{huang2017electro, cao2014structure, gu2001prolonged} 
that alter the amino acid sequence. 
Both strategies present challenges mainly associated 
with the capability of the biological pore to properly self-assemble into the lipid membrane and form a well 
defined structure even under extreme pH conditions and
after mutation of exposed amino acids.
Nevertheless, some pores are remarkably robust. 
For instance,
$\alpha$-Hemolysin forms stable pores 
from $\mathrm{pH}~2.8$ to $\mathrm{pH}~11$~\cite{
		gu2001prolonged,
		betermier2020single,
		asandei2016electroosmotic}
while in the constriction of the CsgG nanopore, the central amino acid (Asn-70) 
can be mutated in any other of the 19 standard proteinogenic 
amino acids~\cite{cao2014structure}
without altering the stability of the assembled nonameric channel.

Solid state nanopores usually present
a more uniform surface charge 
than biological pores.
The charge can be either 
positive (\textit{e.g.} $\mathrm{Al_2O_3, ZnO}$) or
negative (\textit{e.g.} $\mathrm{SiN, SiO_2}$) 
depending on the interfacial properties of the material
and the fabrication process~\cite{xue2020solid}.
The charge can be tuned by changing the salt concentration~\cite{wanunu2007chemically,lin2021surface}
and also the 
pH~\cite{chen2004atomic,hoogerheide2009probing}.
For example, the surface charge of $\mathrm{SiN}$ nanopores 
ranges from
$0.0027~\mathrm{C/m^2}$ at $\mathrm{pH}~1.2$ 
to
$-0.2~\mathrm{C/m^2}$ at $\mathrm{pH}~11$~\cite{lin2021surface}.
In addition, multiple coating techniques are available to functionalize
solid membranes 
and transfer to synthetic nanopores some properties of biological 
nanopores~\cite{eggenberger2019surface,lepoitevin2017functionalization},
including: deposition from gas phase, 
surfactant adsorption, physisorption, 
monolayer and layer-by-layer self-assembly, and silanization.
The charge of the functional groups, see Fig.~\ref{fig:mechanism}c, 
can be further tuned by changing the pH, the electrolyte type and its
concentration~\cite{yameen2009single,anderson2013ph,ma2019nanopore,karmi2020durable,tanimoto2021selective},
allowing one to completely neutralize or invert 
the pristine pore charge and, hence, the EOF direction. 

As a final comment, it is worth noting that
a fixed surface charge is not the only way to achieve 
such a kind of intrinsic selectivity, {\sl i.e.} 
an accumulation of positive or negative ions 
in the absence of any external perturbation.
Indeed, intrinsic net charge accumulation in confined 
geometries even with a 
zero surface charge of the solid was
observed in atomistic 
simulations~\cite{kim2009high,di2021geometrically}.
In brief, in real electrolytes
an equilibrium charge layering spontaneously arises at the solid-liquid interface
due to the different sizes and solvation energies of cations and anions. 
Moreover, the preferential orientation of water molecules 
at the walls results in net interfacial dipoles.
The presence of interfacial dipoles generates an intrinsic 
polarization of the membrane, resulting in an effective surface potential.

\paragraph*{Voltage gating
\footnote{
	Voltage gated nanopores described in this review should not be confused with \textit{voltage gated ion channels}~\cite{armstrong1998voltage}. The main difference is that in voltage gated ion channels the transmembrane voltage induces conformational changes which influence their transport properties. In the voltage gated nanopores described here, an additional external voltage is applied to the membrane, providing independent control of transmembrane voltage and surface potential at the membrane.
	}.}
A second approach to generate
charge accumulation
is to embed a gate electrode electrically connected to the membrane, Fig.~\ref{fig:mechanism}d.
In that way, the $\zeta$-potential is actively modulated 
by controlling the voltage of the gate, $\Delta V_G$~\cite{guan2014voltage,ren2017nanopore}.
In essence, the strategy is similar 
to the MOSFETs electron/hole population regulation 
in silicon conductive channels used in electronics~\cite{sedra2010microelectronic}.
Gate electrodes are usually fabricated with thin metal 
films~\cite{kalman2009control,van2018ion,bai2014fabrication,nam2009ionic}
(see for example the TiN metal layers in Fig.~\ref{fig:mechanism}e), conductive polymers~\cite{perez2017all}
or single-layer graphene sheets~\cite{cantley2019voltage}
(Fig.~\ref{fig:mechanism}f).
The strenuous fabrication process
(including a sacrificial layer, bonding, 
high temperatures for the film deposition, 
electron beam lithography or atomic layer deposition) 
often limits the resolution of experimental designs 
to channels with diameters $10\--100$~nm.
However, 2D sub-nanometric fluidic confinement 
is possible with voltage-gated materials
fabricated with multiple layers of packed, 
electrically conductive, nanosheets 
(such as graphene~\cite{cheng2018low} and MXene~\cite{wang2019voltage}).

\paragraph*{Induced charge.}  
The third mechanism exploits the externally imposed voltage drop between 
the two compartments to also induce charge accumulation in the pore.
Indeed when an electric field is applied, ions migrate
towards or away from the membrane, and can, under certain conditions, 
generate a local accumulation of net charge~\cite{squires2009induced}. 
This net charge is temporary and is released as the voltage drop is removed.
The region in which charge accumulates has a thickness of the order 
of the Debye length and it is named Induced Debye Layer, 
IDL~\cite{lauger1967electrical,bazant2010induced,di2021geometrically}.
If charge accumulates inside the pore region, 
the same electric field can also induce an EOF.

The origin of the charge accumulation in the IDL and thus of the EOF 
is related to the normal and tangential components of 
the local electric field at the membrane/liquid interface which, 
in turn, depend on the pore geometry and on the fluid and 
membrane permittivities.
This can be seen considering the jump conditions 
derived from Gauss's law and the fact that the electric field 
is irrotational for two materials with different 
dielectric properties
\begin{align}
\label{eq:jump1}
\left(\varepsilon_F\boldsymbol{E}_F-\varepsilon_M\boldsymbol{E}_M\right)\cdot\boldsymbol{\hat{n}}=q_w \;,\\
\label{eq:jump2}
\left(\boldsymbol{E}_M-\boldsymbol{E}_F\right)\times\boldsymbol{\hat{n}}=0\;,
\end{align}
where $\varepsilon_F$ and $\varepsilon_M$ are the electric 
permittivities of the fluid and the membrane, 
$\boldsymbol{E}_F$ and $\boldsymbol{E}_M$ are the electric fields at the boundary, 
respectively on the fluid and membrane side, 
$q_w$ is the surface charge and $\boldsymbol{\hat{n}}$ is the outward normal to the membrane.
The normal component of $\boldsymbol{E}_F$ drives the formation of the IDL, 
see Fig.~\ref{fig:mechanism}g, left side.
The tangential components of $\boldsymbol{E}_F$
will move these accumulated charges along the channel wall, 
generating the EOF, usually called 
induced-charge electroosmosis (ICEO)~\cite{squires2009induced}.
It is worth noting that an asymmetry is needed to generate 
a net EOF, since a perfectly symmetric system would generate
perfectly symmetric ion flows and no net charge accumulation in the pore, 
see the cylindrical pore example in Fig.~\ref{fig:mechanism}g, right side.

In a lumped-parameter model,
it is possible to describe the membrane as a capacitor able to accumulate 
and release charge in response to the external voltage drop,
$q_a = C \Delta V$, with $q_a$ the {\sl local} accumulated charge (IDL) 
and $C$ the {\sl local} membrane capacitance, Fig.~\ref{fig:mechanism}g.
Considering the simple case of a planar solid membrane of thickness $h$ which is
immersed in an electrolyte solution,
the capacitance per unit surface can be approximated by
$
	C\simeq 
	\varepsilon_0 \varepsilon_M h^{-1}
$
~\cite{lauger1967electrical,di2021geometrically}. 
Induced charge accumulation hence becomes particularly noticeable 
at sharp corners and in tiny nanopores,
where the thickness of the solid substrate becomes very 
small~\cite{yao2020induced, di2021geometrically}.
It is worth noting that the existence of a non-zero $\boldsymbol{E}_M$,
Eq.\eqref{eq:jump1}, is needed for the formation of the IDL~\cite{lauger1967electrical,di2021geometrically}, 
and hence simplified models that for $\varepsilon_M \ll \varepsilon_F$ neglect ${\bf E}_M$
are not able to describe IDL and ICEO.

Such nanoscale ICEO has recently been investigated numerically. 
As a first example, in Fig.~\ref{fig:mechanism}h 
single-polarity ions in the Debye layer and their
counterions massively accumulate near the edge of a conical nanopore. This accumulation 
results in the formation of electroosmotic vortices~\cite{yao2020induced}.
Another example is reported in Fig.~\ref{fig:mechanism}i, 
where geometrical symmetry is broken by a surrounding cavity 
outside the pore and EOF is achieved in the absence of fixed charges ($q_w=0$).
This approach induces ion selectivity
without altering the pore shape, surface charge
or chemistry and, consequently, opens new possibilities
for more flexible designs of selective nanoporous membranes~\cite{di2021geometrically}.

Finally, an intriguing and important feature of ICEO is that the induced flow 
depends quadratically on the applied voltage 
$\Delta V$~\cite{squires2009induced,di2021geometrically}.
Indeed, 
the EOF scales roughly as $q_a \Delta V$,
where $q_a$ is also proportional to $\Delta V$.
This quadratic dependence
results in unidirectional EOF, \textit{i.e.}
the direction of the EOF is always the same, 
even when the applied voltage drop is inverted, see Fig.~\ref{fig:mechanism}i.
Hence, a net fluid flow can be generated using both AC or DC fields.

\subsection{Modeling and computational challenges}

The diversity of contexts in which electroosmosis arises hints at the potential hardships to properly model such flows. Indeed, a reliable description of EOF is challenging for several reasons that we recapitulate below. 

First,
it requires to describe precisely numerous forces.  
In fact, it should model the hydrodynamic 
transport of the fluid and the ionic species dissolved therein,
the electrostatic interactions among the different charged species 
and with the solid surfaces, 
the polarization of the membrane and the effect of external forcings. 
In addition,
 the presence of particles such as colloids or proteins, 
which typically present peculiar charge distributions, 
requires additional modeling of their interactions 
with the electrolyte solution. 

A further challenge comes from the wide range of spatial and temporal scales that are relevant to the phenomenon. 
To illustrate the diversity of scales at play, we consider a typical nanoporous
sensing device. A voltage drop $\Delta V$ is applied between the two reservoirs, see 
Fig~\ref{fig:mechanism}a, and we probe
the role of electroosmosis for the capture
of an analyte (\textit{e.g.} a protein, a nucleic acid, 
a pollutant) by the pore.
The pore constriction is usually of the order of a few nanometers. 
The applied voltage drop $\Delta V$ results in 
a funnel-like electric field outside the pore, decaying slowly as $1/r^2$ with $r$ the distance to the pore~\cite{wanunu2010electrostatic,chinappi2015nanopore}.
The electric field is therefore considerable even decades of nanometers away from the
pore entrance. 
A reliable model of the flow in this external {\sl capture} region, on scales much broader than the nanoscale pore, is crucial 
to determine the motion of the analyte from the bulk to the
pore entrance. 
Along with this diversity of spatial scales, very different time scales are at play. The motion of analytes needs to be resolved inside the constriction and is quite fast as it only occurs over a short spatial range. 
However capture events need to be resolved over long time scales as they are usually rare events.

Finally, as the system is usually extremely confined -- such as in nanopores --
thermal fluctuations have to be taken 
into account. Thermal vibrations of the pore's structure affect the motion of analytes~\cite{marbach2018transport,ma2015water}. 
The number of analytes within the pore is also strongly fluctuating, as there are only a few particles within at a time~\cite{gravelle2019adsorption,thorneywork2020direct,marbach2021intrinsic}. 
Moreover, the analyte undergoes Brownian motion,
so the interest in not on the analysis of a single capture event, but on
a statistical description of 
the capture~\cite{chinappi2020analytical,wong2007polymer,grosberg2010dna}.

There is no single computational method
that is
able to handle all the aforementioned physical features -- variety of forces, space and time scales, and intrinsic fluctuations -- 
with the currently available high performance computational resources.
Selectivity for specific ionic species by the nanopore constriction
is ruled by electrochemical interactions that occur at 
nanometer scales thus calling for an atomistic 
description~\cite{aksimentiev2005imaging,bhattacharya2011rectification,bonome2017electroosmotic}, 
which is discussed in Section~\ref{sec:md}.
However,
computational requirements make atomistic models unsuited for the modeling of the 
capture region, and, more importantly, prevent the implementation 
of efficient, computer-assisted, design strategies that usually require 
the exploration of a wide number of different 
operating conditions.
Indeed, even on supercomputers, atomistic simulations hardly reach a microsecond/day, while typical capture and translocation time scales range from milliseconds 
to seconds~\cite{boukhet2016probing,soskine2012engineered,schmid2021nanopore,asandei2016electroosmotic}.
Standard continuum models, discussed in Section~\ref{sec:continuum}, are computationally less demanding 
and they enable the description of long time scales but, on the one hand, 
they often require {\sl ad hoc} nanoscale corrections, and, on
the other hand, they do not include thermal fluctuations. 
Mesoscale models, discussed in Section~\ref{sec:meso}, attempt to bridge the gap between continuum
and atomistic descriptions, but often require external information 
(\textit{e.g.} coarse-grained modeling of chemical interactions) that may 
need to be finely tuned to obtain quantitative results.
In the following
we briefly review these different approaches with the aim to help 
researchers select the technique that better reflects the levels of accuracy and 
approximations suitable to answer a specific question. 
Finally, some of the most challenging applications related to EOF in nanopores are reported 
in Section~\ref{sec:techno}.

%% file: continuum.tex
\section{Continuum methods}
\label{sec:continuum}

We start by reviewing how continuum models may be used to explore EOF, including a careful explanation of specific modeling assumptions that have to be made. We discuss representative examples. Finally we explore the limitations of such continuum approaches. 

Continuum models rely on a set of equations: the continuity equation for each species, 
the momentum and mass balance for the fluid, 
and the Poisson equation for electrostatics,

\begin{align}
\label{eq:continuity}
\frac{\partial c_\alpha}{\partial t}+\boldsymbol{\nabla}\cdot\left(c_\alpha\boldsymbol{u}+\boldsymbol{j}_\alpha\right)=0\;,\\
\label{eq:momentum}
	\rho\frac{\partial\boldsymbol{u}}{\partial{t}}+\rho\boldsymbol{u}\cdot\boldsymbol{\nabla}\boldsymbol{u}=\boldsymbol{\nabla}\cdot\boldsymbol{\Sigma}-\rho_e\boldsymbol{\nabla}\Phi\;,\\
\label{eq:incompressible}
\boldsymbol{\nabla}\cdot\boldsymbol{u}=0\;,\\
\label{eq:poisson}
\boldsymbol{\nabla}\cdot\left(\varepsilon\boldsymbol{\nabla}\Phi\right)=-\rho_{e}\;,
\end{align}
whose derivation can be found in standard microfluidics 
textbooks~\cite{theoretical_microfluidics,micronanofluid}.
For each dissolved species $\alpha$, Eq.~\eqref{eq:continuity} describes the time evolution of the number density $c_\alpha$.
The flux has two contributions, a convective flux $c_\alpha\boldsymbol{u}$, where $\boldsymbol{u}$ is the fluid velocity, 
and a nonconvective flux $\boldsymbol{j}_\alpha$ which has to be specified via a constitutive relation.
Eqs.~\eqref{eq:momentum}-\eqref{eq:incompressible} describe momentum and mass balance of an incompressible flow.
Here, $\rho$ is the (constant) fluid density, $\boldsymbol{\Sigma}$ is the stress tensor to be specified by a constitutive relation, $\Phi$ is the electrostatic potential and $\rho_e$ is electric charge density, 
which is expressed in terms of the ionic species concentration $c_\alpha$ by
$$
	\rho_e
        =
        \sum\limits_{\alpha=1}^{N_s} c_\alpha q Z_{\alpha}\;,
$$
where $N_s$ is the total number of ionic species, $q Z_\alpha$ is the charge of species $\alpha$.
Most fluids relevant to EOF are Newtonian fluids, for which the stress tensor reads
\begin{equation}
 \boldsymbol{\Sigma}
 =
	-p\boldsymbol{I}+2\eta\boldsymbol{\sigma}\;,
\label{eq:newt}
\end{equation}
where $p$ is the pressure field, $\boldsymbol{I}$ is the identity tensor, $\eta$ is the viscosity of the fluid 
and $\boldsymbol{\sigma}=(\boldsymbol{\nabla u}+\left(\boldsymbol{\nabla u}\right)^T)/2$ is the strain rate tensor. 
When Eq.~\eqref{eq:newt} is used, 
the momentum and mass balance,  Eqs.~(\ref{eq:momentum}-\ref{eq:incompressible}), are referred to 
as the Navier-Stokes equations.
Complex fluids
require more specific constitutive relations instead of Eq.~\eqref{eq:newt}, 
see, \textit{e.g.}~\cite{mukherjee2017ion}
 where EOF in 
viscoelastic fluids is discussed.
At sufficiently small scales, typical of nanopores, inertial and nonlinear terms, 
\textit{i.e.} the left hand side of Eq.~\eqref{eq:momentum}, 
may usually be neglected\footnote{
Except around a non-zero, large, average flow, where the left hand side has to be linearized. 
}, leading to the Stokes equation~\cite{microhydrodynamics}. 
In the Navier-Stokes equations, Eq.~\eqref{eq:momentum} the term $-\rho_e\boldsymbol{\nabla}\Phi$ 
is crucial for the description of EOF: it drives solvent flow where charge is accumulated.
Finally, Eq.~\eqref{eq:poisson} is the Poisson equation, derived from Gauss's law in a medium of 
permittivity $\varepsilon = \varepsilon_0 \varepsilon_r$. 
The set of equations~(\ref{eq:continuity}-\ref{eq:poisson}) is closed once the 
flux of ionic species $\boldsymbol{j}_\alpha$ is specified. 
Within linear response, considering standard Fickian diffusion and electrophoretic motion of the ions
\begin{equation}
\label{eq:nernstplanck}
\boldsymbol{j}_\alpha=-D_\alpha\left(\boldsymbol{\nabla} c_\alpha+c_\alpha\frac{qZ_\alpha}{k_BT}\boldsymbol{\nabla}\Phi\right)\;,
\end{equation}
Eq.~\eqref{eq:continuity} becomes the Nernst-Planck equation, which is widely used in the EOF literature. 
Eq.~\eqref{eq:nernstplanck} embeds several assumptions on the nature of the solution, which are briefly discussed in section~\ref{sec:limitations}.

The set of Eqs.~(\ref{eq:continuity}-\ref{eq:poisson}) for a Newtonian fluid Eq.~\eqref{eq:newt} and a standard 
flux given by Eq.~\eqref{eq:nernstplanck} are known as the Poisson-Nernst-Planck-Navier-Stokes equations (PNP-NS), and, once proper boundary conditions are imposed, can be solved self-consistently via computational methods such as finite elements (FEM) or finite volumes (FVM)~\cite{prohl2010convergent,sherwood2014electrically}.

\subsection{Boundary conditions}

When modeling EOF in nanopores, care should be given to
the selection of boundary conditions.
It is convenient to divide boundaries in two groups: reservoir boundaries 
and membrane boundaries. 
In general, the domains on which the Poisson equation and the transport equations are solved differ.
Transport equations are solved only in the fluid domain
while the Poisson equation needs to be 
solved both in the fluid and in the membrane,
in particular when the electric field inside the membrane 
is relevant, as is the case for induced charged EOF as discussed in section~\ref{sec:3routes}, see Fig.~\ref{fig:mechanism}~III.
\\

{\bf Transport equations.} For the ionic transport equations,
it is reasonable to consider that far from the pore at both sides
two reservoirs are present, each with a fixed concentration.
This translates to the condition
$c_\alpha=c_{\alpha,cis}$ and $c_\alpha=c_{\alpha,trans}$ at the boundaries separating the system from the two reservoirs, see Fig.~\ref{fig:cont}a (top).
If the ions cannot penetrate the membrane the impermeability
condition $\boldsymbol{\hat{n}}\cdot\boldsymbol{j}_\alpha=0$ holds at the fluid-membrane interface,
where $\boldsymbol{\hat{n}}$ is the outward normal to the membrane surface,
see \textit{e.g.} Fig.~\ref{fig:cont}a (top).
For what concerns fluid transport, usually it can be assumed that 
no mechanically induced flow is present.
In this case, a zero stress condition can be used at the reservoir boundary
$\boldsymbol{\hat{n}}\cdot\boldsymbol{\Sigma}=0$. 
In the case of pressure driven flow,  
the pressure at the inflow and outflow boundaries needs to be specified~\cite{cho2012characteristics}.
As for the fluid-membrane interface, the impermeability condition 
applies to the normal component of the velocity 
$\boldsymbol{u}\cdot\boldsymbol{\hat{n}} = 0$. 
The boundary condition for the tangential component of the velocity requires some assumptions
on the nature of the fluid-membrane interactions. 
If these are such that the fluid in contact with the membrane has zero velocity, 
a no-slip condition $\boldsymbol{u}=0$ has to be used.
However, especially inside the nanopore, 
the no-slip condition may fail to represent the fluid-membrane 
interaction and can be substituted by the more general 
Navier slip condition, in which the stress exerted by the membrane on the fluid is proportional to the velocity of the fluid at the interface~\cite{lauga2007microfluidics,collis2021measurement},
$\left(2b \,\boldsymbol{\hat{n}}\cdot\boldsymbol{\sigma}-\boldsymbol{u}\right)\cdot\left(\boldsymbol{I}-\boldsymbol{\hat{n}}\otimes\boldsymbol{\hat{n}}\right)=0$  (where $b$ is usually a scalar except for anisotropic surfaces~\cite{bazant2008tensorial}).
The slip length $b$ is a parameter which quantifies the motion of the fluid with respect to the wall at the boundary~\cite{sendner2009interfacial} and characterizes the liquid/solid interactions.
For example slippage increases with hydrophobicity~\cite{huang2008water,chinappi2010intrinsic,bakli2015slippery}. 
In practical cases of hydrophilic or slightly 
hydrophobic membranes, $b$ is usually in the nanometer range, and is therefore mostly relevant in the pore.
Yet even a subnanometer slip length
may strongly affect the EOF intensity~\cite{joly2006liquid}.
\\

{\bf Poisson's equation.} 
The membrane and the fluid have in general different dielectric properties and often a surface charge $q_w$ is present at the fluid-membrane interface. 
Hence, the electric field can be discontinuous at the interface. This
requires to solve Poisson's equation separately in two subdomains, the fluid 
and the membrane subdomains, introducing an internal boundary 
(\textit{i.e.} the fluid-membrane interface where a matching 
condition should be imposed) and an external
boundary (\textit{i.e.} the whole domain boundary).
At the interface between the fluid subdomain and the reservoirs, a voltage drop can be imposed by setting
$\Phi=0$ at one reservoir and $\Phi=\Delta V$ at the other, see Fig~\ref{fig:cont}a. 
At the external boundary of the membrane, a vanishing normal component of the electric field
can be used 
($\boldsymbol{\hat n} \cdot {\bf E} = 0$)
-- assuming that the domain is sufficiently 
large to consider the nanopore far enough, see the small brown external boundary of the membrane in Fig.~\ref{fig:cont}a.
For the internal boundary, 
the appropriate equations are given by the already mentioned jump conditions 
derived from Gauss's law and the irrotationality of the electric field, see Eqs.~(\ref{eq:jump1}-\ref{eq:jump2}).
A surface charge in the membrane can be represented either by explicitly setting $q_w$ in the jump condition of Eq.~\eqref{eq:jump1} or by adopting a sufficiently thin slice of volumetric charge density inside the membrane close to its surface, as for instance in~\cite{melnikov2017electro}.
%

\subsection{PNP-NS model to study electroosmotic flows}

The PNP-NS model has been widely used to model EOF. 
Under the Debye-H\"uckel approximation (\textit{i.e.} small $\zeta$-potential), the set of Eqs.~(\ref{eq:continuity}-\ref{eq:poisson}) can be solved with semi-analytical perturbative approaches in case of channels of smoothly varying section and sufficiently low external voltages.
These approaches have been used to capture important qualitative features, such as the presence of recirculating regions in which the flow direction is opposite as compared to the average volume flow~\cite{malgaretti2014entropic}, as confirmed also via molecular dynamics simulations~\cite{chinappi2018charge}.
These approaches can also be used to characterize the linear response of such channels in terms of fluxes (electric, concentration and mass) under different external stimuli~\cite{malgaretti2019driving}.

\begin{figure}
	\center
	\includegraphics[width=\linewidth]{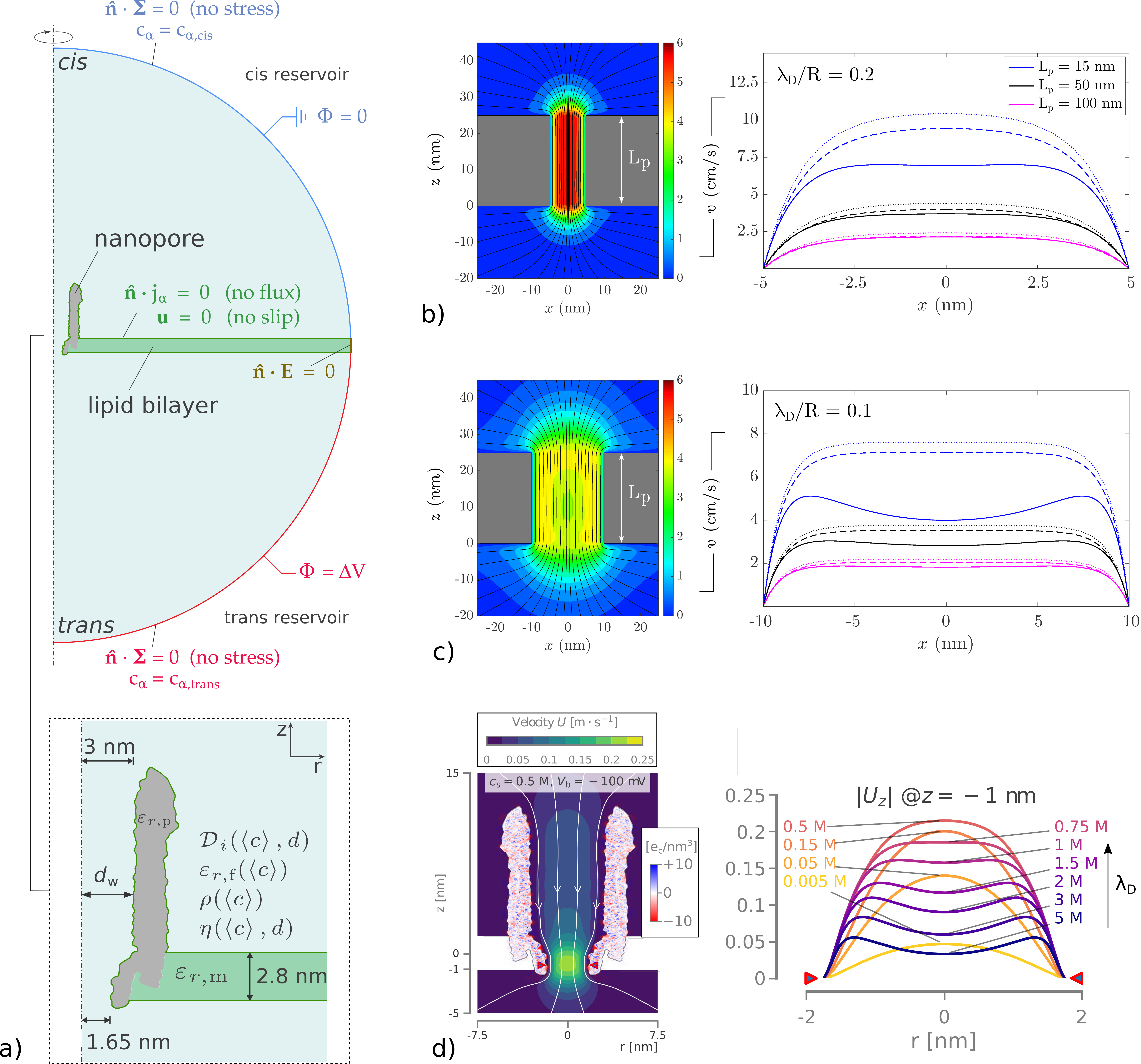}
	\caption{
		\textbf{Examples of EOF solved within the PNP-NS framework.}
		Panels a and d are adapted from Willems et al.~\cite{willems2020accurate},
		while panels b and c are adapted from Melnikov et al.~\cite{melnikov2017electro}.
		\textbf{Boundary conditions. a)}  Model for a typical nanopore geometry 
		with the different components of the PNP-NS equations and boundary conditions.
		Here, dielectric constant $\varepsilon$, viscosity $\eta$, diffusivity $D$ and 
           	fluid density $\rho$
		depend on the local ion concentration, 
	        while $\eta$ and $D$ also depend  on the distance to the 
		solid boundary. 
		\textbf{Effect of pore size. b)} Velocity field and axial velocity profile inside a 
		cylindrical nanopore with radius $5$~nm.
		The surface charge of the membrane and the voltage drop $\Delta V$ 
		applied via boundary conditions give rise to EOF.
		Solid lines on the velocity profile represent the 
		result of PNP-NS numerical simulations,
		while dashed and dotted lines show the analytical solution
		for the linearized Debye-H\"{u}ckel theory for an infinitely long pore,
		using two different expressions for the $\zeta$-potential.
		\textbf{c)} Same as b), considering a nanopore with larger radius, $10$~nm.
		\textbf{Effect of concentration. d)} Electroosmotic velocity field inside 
		a biological nanopore (ClyA) and axial velocity profile at the constriction.
		The biological nanopore
		is modeled as a fixed spatial charge density.
		The geometry of the nanopore, extracted from its molecular structure,
		is used as a boundary for the fluid domain.
		The  velocity profile is shown for different 
		ion concentrations (hence, different $\lambda_D$).
		}
	\label{fig:cont}
\end{figure}

In the case of high $\zeta$-potentials and external fields, the solution of Eqs.~(\ref{eq:continuity}-\ref{eq:poisson}) may quantitatively differ from the one obtained with linearized approaches.
Moreover, when the channel section varies abruptly, as in many nanopore applications, such approaches may fail to properly describe the ionic distributions~\cite{malgaretti2014entropic}. 
In these cases, the numerical solution of the complete non linear PNP-NS system may be required.
For example, Melnikov et al.~\cite{melnikov2017electro},
showed that, even for simple systems such as charged cylindrical nanopores, 
finite pore length leads to substantial differences 
with respect to analytic results 
regarding infinitely long channels of constant section. 
The velocity field of the nonlinear PNP-NS for a charged finite cylinder for two different pore radii is reported in Fig.~\ref{fig:cont}b and c, left side.
The axial velocity profiles inside the pore 
 (full lines) are compared with analytic predictions 
for infinite cylinders (dashed and dotted) in Fig.~\ref{fig:cont}b and c, right side.
The PNP-NS solution differs quantitatively from the analytic result, 
especially for shorter pores. 
This can be ascribed to the assumption of infinite channel length which neglects pore entrance effects.

Furthermore, interesting qualitative features of the velocity profile inside pores may be revealed with continuum models -- and differ from analytic predictions for infinitely long cylinders (Fig.~\ref{fig:cont}b and c). 
Due to the fluid incompressibility, a pressure difference builds up between the two ends of the pore. 
This pressure drop generates a force density which opposes the electroosmotic flow.
If the pore is sufficiently short, this force density may overcome the electrostatic force density contribution, 
inducing a change in the concavity of the velocity profile.
Since the electroosmotic driving force is mainly located in the Debye layer, this effect depends 
on the ratio of the Debye length $\lambda_D$ to the pore radius.
With a short Debye length with respect to the pore radius, the flow is surface-driven 
and hence the force density generated by the pressure drop induces a relative minimum 
at the center of the pore, where the electrostatic contribution vanishes.
In a system in which the Debye length is comparable to the pore radius, the flow is bulk-driven and the force density induced by the pressure drop is counterbalanced by the electrostatic force density, now relevant also in the center of the pore, resulting in a maximum of the velocity profile at the center~\cite{melnikov2017electro}.

More complex pores have also been explored by continuum methods. 
Biological nanopores have an extremely complex geometry and charge distribution which makes them difficult to simulate in a continuum framework.
Still, continuum models of biological pores can be built and investigated, 
for example Willems et al.~\cite{willems2020accurate} 
investigated a continuum model for the geometry and charge of Cytolysin A (ClyA), see Fig.~\ref{fig:cont}d, a toxic protein produced by \textit{E. Coli}.
An extended PNP-NS approach in which steric effects 
have been added to the constitutive relation Eq.~\eqref{eq:nernstplanck} for the ionic flux~\cite{lu2011poisson}
was used to study the transport of 
ions and water through the pore.
Properties such as ionic mobilities, electric permittivity, 
fluid density and viscosity were 
modeled as dependent on the local ionic concentration and on the 
distance to the liquid-solid interface, see Fig.~\ref{fig:cont}a (bottom). 
The resulting velocity field is shown in Fig.~\ref{fig:cont}d (left side), 
with axial velocity profiles computed at the pore constriction 
for different salt concentrations shown in Fig.~\ref{fig:cont}d (right side).
Interestingly, the change in the velocity profile with increasing salt concentration 
(and hence decreasing Debye length $\lambda_D$) 
shows a minimum in the center of the pore similar 
to the one observed for cylindrical channels, see Fig.~\ref{fig:cont}c. This
suggests both that finite pore length effects are important in biological pores, 
and that a key role is played by the ratio 
between $\lambda_D$ and the pore size in determining the presence of such effects.

In addition to the aforementioned examples, the PNP-NS model has been widely used to study several systems and setups such as induced charge electroosmosis~\cite{yao2020induced,davidson2014chaotic}, EOF rectification~\cite{laohakunakorn2015electroosmotic}, EOF reversal in a glass nanopore~\cite{laohakunakorn2015reversal} and the effect of EOF on ionic current rectification~\cite{ai2010effects}.
In all these cases, nonlinear effects due to charge redistribution under the applied voltage are crucial, and cannot be captured by analytical approaches based on the linearized Poisson-Boltzmann equation. 
As previously reported, semi-analytical perturbative approaches~\cite{malgaretti2014entropic,malgaretti2019driving} may be used to study EOF in smoothly varying channels.
Such semi-analytical approaches are much more computationally efficient when compared to PNP-NS and can therefore prove useful when a fast screening of the EOF in several operating conditions is needed, \textit{e.g.} for design purposes.
Similarly, analytical models for electric conductance including also the effect 
of liquid slippage at the wall~\cite{manghi2021ionic,balme2015ionic},
 and EOF models for finite cylindrical pores~\cite{sherwood2014electroosmosis}
have been developed. Although these models are often limited
to simplified geometries ({\sl e.g.} cylindrical pores), they can still provide preliminary indication 
of the magnitude of the currents.

\subsection{Some limitations of the PNP-NS model}\label{sec:limitations}

Despite the complexity and variety of systems which can be studied, the PNP-NS model suffers from some limitations, and requires variations or entirely different approaches to tackle specific problems. We explore 3 major limitations below.

{\bf Electrolyte model.} The PNP-NS equations rely on a model for the ion flux, Eq.~\eqref{eq:nernstplanck}, which is based on several assumptions.
In particular, the chemical potential of the solvent is not taken into account,
while the chemical potential for solutes is assumed to be well described by an ideal (dilute) solution approximation, $\mu_\alpha=k_BT\ln\left(c_\alpha/c_0\right)+Z_\alpha q\Phi+\tilde{\mu}_\alpha$, where $c_0$ is a reference concentration, and $\tilde{\mu}_\alpha$ is a constant.
Eq.~\eqref{eq:nernstplanck} also assumes a diagonal diffusivity, meaning that the motion of the different species is uncorrelated.
For a detailed discussion of other underlying assumptions, we refer the reader to Dreyer \textit{et al.}~\cite{dreyer2013overcoming}. 
All these assumptions limit the scope of the results. For example, at high concentrations the dilute approximation breaks down and an alternative model for the chemical potential has to be used~\cite{marbach2017osmotic}. More complex models for the chemical potential can be formulated to obtain expressions for the flux which take into account specific features of the solution, such as steric and solvation effects~\cite{noneq_thermodynamics,fuhrmann2015comparison,bandopadhyay2013ionic,chaudhry2014stabilized,qiao2014ionic,siddiqua2017modified}.
Taking into account steric effects leads to reduced charge accumulation near the charged surfaces, mitigating the highly nonlinear effects arising when high $\zeta$-potentials are involved~\cite{kilic2007steric1,kilic2007steric2}.
This overestimation of the charge density may significantly affect the predicted EOF~\cite{fuhrmann2019induced}. 

{\bf Confinement.}
Beyond modeling assumptions for the bulk electrolyte, when dealing with nanopores, the behavior of the solution is dramatically affected by the extreme confinement in the pore region.
In fact, from a few tens of nanometers and below, the growing importance of surfaces triggers 
a diversity of surprising effects,
challenging the continuous description of hydrodynamics Eqs.~(\ref{eq:momentum}-\ref{eq:incompressible}). 
First, water transport becomes strongly affected by the interactions with the pore 
wall material
inducing an effective slip of water, discussed above, which can enhance the EOF~\cite{joly2006liquid}.
Slippage at the interface may also be dependent on the structure of the confined ions, 
especially in subnanometer pores~\cite{mouterde2019molecular}. 
Furthermore, at subnanometer confinement water reorganizes in layers, 
strongly suppressing dielectric permittivity~\cite{schlaich2016water,fumagalli2018anomalously}. 
An effective dielectric permittitivity may be used in Poisson's equation Eq.~\eqref{eq:poisson}, 
see \textit{e.g}~\cite{kavokine2019ionic}. 
The mobility of ions depends on their distance to the wall, which can be modeled by \textit{e.g.} 
phenomenological diffusion coefficients for the ions~\cite{willems2020accurate}. 
Finally, the interplay between ionic and water transport at interfaces modifies currents: 
for example phenomena akin to passive voltage gating have been observed in confinements smaller 
than 2~nm, where ionic mobility under applied pressure depends on the applied voltage 
drop~\cite{mouterde2019molecular}. 
Naturally, it is expected that in similar osmotic-like transport, such as EOF, such curious coupling would also arise. 

Such effects undoubtedly call for atomistic models, that we discuss in section~\ref{sec:md}. 
Note that, at small confinements, standard atomistic models such as molecular dynamics, 
may still fail to reproduce quantitative agreement between measurements and simulations. 
In fact, electronic interactions are at play and only \textit{ab initio} 
descriptions can provide accurate 
models~\cite{sokoloff2018enhancement,kavokine2021fluctuation}.

{\bf Fluctuations.}
As a final remark, the PNP-NS model relies on a mean field approach, in which the fluctuations with respect to 
the average value of the fields (\textit{i.e.} velocity, concentration, voltage, surface position) are not taken into account.
Such fluctuations are present at all scales but their importance increases for the smallest nanopore systems~\cite{siwy2002origin,secchi2016scaling,knowles2021current}. 
The presence of thermal fluctuations affects the fluid velocity both directly due to thermal agitation of the solvent molecules, and indirectly, since fluctuations in the ionic distribution affect the driving electroosmotic force. Hence, continuum descriptions may fail to be either quantitatively or even qualitatively correct in very narrow nanopores.
We refer the reader to Ref.~\cite{kavokine2021fluids}
for further insight on the breakdown of continuous equations in confinement.

%% file: atomistic.tex
\section{Atomistic description}
\label{sec:md}
\subsection{Challenges in the numerical implementation}
Atomistic simulations have been widely used to study transport phenomena at the 
nanoscale~\cite{murad1998molecular,
joly2006liquid,aksimentiev2005imaging,
zhou2020molecular,di2021geometrically,
aksimentiev2004microscopic,hummer2001water,zhu2004collective,
shankla2019step,bonome2015multistep,
farimaniidentification,shankla2020molecular,hu2012ion,
chinappi2018charge,qiao2003ion,
qiao2003atypical,freund2002electro,comer2011modeling,shahmardi2021effects,
huang2008water,chinappi2010intrinsic}.
The setup is straightforward: 
each atom is described as a classical
material point of specified mass and charge.
The material points interact via 
conservative forces and the system evolves in time according to
Newton's second law.
In the literature, this approach is often
referred to as all-atom Molecular Dynamics 
to be distinguished from coarse-grained 
methods~\cite{marrink2007martini,clementi2008coarse,maffeo2014coarse}
where material points do not necessarily correspond
to single atoms. In the following, for simplicity 
we use Molecular Dynamics (MD) to refer
to all-atom approaches.

The time evolution of such a mechanical system with $N$ particles
in a limited volume $V$ interacting via conservative 
forces (such that the total energy $E$ is constant) 
corresponds to a thermodynamically 
isolated system, that samples 
the microcanonical ensemble (NVE).
In practical applications, NVE systems 
are quite rare and, consequently, 
MD approaches were complemented 
by several, now standard, tools.
Modifying the dynamics
allows one to sample other statistical mechanical
ensembles such as the canonical (NVT, with $T$ temperature)
and the isobaric-isotermic (NPT, with $P$ pressure) ensembles.
We refer the reader to classical 
resources~\cite{tuckerman2010statistical,allen1989computer,frenkel2001understanding}
for more details. 
We focus here on three specific
aspects that we believe to be especially relevant
(and, in some cases, somewhat overlooked) 
to model EOF across nanopores, 
namely \textit{(i)} what pore model is used, \textit{(ii)} how to acknowledge for large reservoirs and especially how many particles to simulate within the pore in the absence of reservoirs and \textit{(iii)} how to infer transport properties out of equilibrium (in the context of EOF when a voltage drop is applied). 
   \\

{\bf i) Pore model: structure and interaction forces.} For each pore investigated, in MD simulations its structure and the effective interaction forces between the pore atoms and the other species have to be specified . According to the type of pore under scrutiny (biological or artificial), modeling challenges are different. 

A fundamental requirement for 
the simulation of a biological pore 
is the presence of a reliable 
experimentally-determined structure (\textit{i.e.} the conformation of the folded proteins constituting the pore).
Macromolecular structures can be determined 
from protein crystals using a variety of methods, 
including 
X-Ray Diffraction/X-ray crystallography~\cite{mcpherson2014introduction}, 
Cryogenic Electron Microscopy~\cite{thonghin2018cryo} (CryoEM),
Small-angle X-ray scattering~\cite{lipfert2007small}
and Neutron diffraction~\cite{sorensen2018membrane}.
Those structures are typically accessible on 
the Protein Data Bank (PDB~\cite{berman2000protein}),  
while dedicated databases, 
such as OPM~\cite{lomize2012opm} 
and MemProtMD~\cite{stansfeld2015memprotmd}, 
provide spatial arrangements of proteins with respect to 
the lipid bilayer membrane, largely facilitating
the simulation setup. 
Several well-characterized structures
are nowadays available, 
such as the widely studied $\alpha$-Hemolysin ($\alpha$HL~\cite{song1996structure}),
FraC~\cite{tanaka2015structural}  (see also Fig.~\ref{fig:mechanism}-I.b)
and Mspa~\cite{faller2004structure}.

If the structure is not available,
or if it is only partially available,
different 
protein modeling softwares
such as Swiss-Model~\cite{biasini2014swiss}
or MODELLER~\cite{webb2016comparative}
can be used 
to get a complete structure. When the protein portion to be modeled
is located towards the pore lumen
or at the pore entrances, in general such strategies do not guarantee
reliable assessment of ion transport properties (such as selectivity) or EOF. In fact, a slight inaccuracy in the determination of the protein structure could result in a significant modification of the nanoscale confinement. 
Only in a few specific contexts is such modeling reliable: for example in 
CsgG~\cite{goyal2014structural} where the region to be modeled is small and on the exterior of the pore~\cite{di2021geometrically},
or in the Aerolysin pore where the 
 region to be modeled is a small portion
of the extremely stable $\beta-$barrel~\cite{iacovache2016cryo,cao2019single}. 
In both cases, MD simulations were set up 
using standard tools and resulted in stable structures.
Finally, even if a complete structure is available, 
doubts may arise concerning the 
amino acid protonation state, 
in particular if the simulation pH 
is different from the physiological value. 
In that case, common strategies are
to use dedicated bioinformatics tools, 
such as H++~\cite{anandakrishnan2012h++}
or 
PROPKA~\cite{olsson2011propka3}
to predict
the acid dissociation constant $\mathrm{K_a}$
for each titratable residue. $\mathrm{K_a}$ is then used to calculate  
the protonation 
state~\cite{betermier2020single,bonome2017electroosmotic,asandei2016electroosmotic,huang2017electro}. 

In the last decades, reliable atomistic models
for the interaction forces among atoms 
of biological molecules (such as proteins, DNA, lipids)
have been 
developed~\cite{huang2017charmm36m,jorgensen1996development}.
These models, often referred to as {\sl force fields} in
the MD jargon, 
are 
typically 
used for  
the simulation of transport phenomena through biological membranes 
\cite{hammond2021switching,bauer2018mutations,aksimentiev2005imaging,zhou2020molecular,di2019insights}.
For solid state pores, while the structure is inferred by design, 
the reliability of force-fields
is less clear. This is in part due to the diversity of experimental fabrication techniques and material properties for solid-state pores (metallic, non-metallic, semi-conducting) that makes force-field calibrations more challenging. 
Specifically, although classical force fields for common membrane materials are widely used 
(such as for $\mathrm{Si_3 N_4}$~\cite{aksimentiev2004microscopic},
carbon nanotubes 
\cite{hummer2001water,zhu2004collective}
graphene
\cite{shankla2019step,bonome2015multistep}
and 
$\mathrm{MoS_2}$~\cite{farimaniidentification,shankla2020molecular}), 
open issues arise concerning their surface charge
and their effective dielectric constant.
In particular, the determination
of the surface charge for
solid state pores as a function
of the properties of the electrolyte solution 
is {\sl per se} an open issue~\cite{noh2020ion}. Force-fields may however at least be calibrated to reproduce wetting properties (to account properly for fluid-solid interactions) and mechanical properties (to properly reproduce thermal vibrations)~\cite{ma2016fast}.  

Nevertheless, simulations of solid state pores 
may be very useful in electroosmosis 
research since solid state structures
are perfectly suited to 
set up somehow ideal simulations 
aimed at discovering general trends in EOF.
Examples are the 
analysis of the role of electric field
intensity in EOF through a graphene nanopore~\cite{hu2012ion},
the role of 
local electroneutrality breakdown~\cite{noh2020ion}
and eddy formation due to
EOF in varying-section channels
\cite{chinappi2018charge}
and the proof of principle 
of induced charge selectivity in cylindrical neutral
channel discussed in 
Fig.~\ref{fig:mechanism}-III~\cite{di2021geometrically}.
More generally, MD simulations in
simple solid state geometries may be used to
assess the validity limit of continuum
PNP-NS theories and to determine 
peculiar nanoscale effects~\cite{qiao2003ion}.
\\

{\bf ii) Number of molecules within the pore: modeling the effect of reservoirs.} 
The number of molecules within the nanopore is a highly dependent function of the system properties. 
In most experimental settings, the aqueous solution within 
the pore is always in contact with a large 
reservoir where it is possible to control bulk macroscopic conditions,
such as temperature $T$, 
pressure $P$ and salt concentration $c$.
At equilibrium, the number of molecules
that occupy the pore is ruled by chemical
potential equilibrium between the bulk and pore regions.
For pores of size $R$ much larger than the Debye length 
$\lambda_D$ and the molecule size $a$, typically bulk conditions are reproduced inside the pore.
The number of molecules can
be reliably approximated as the 
pore volume times the bulk concentration 
of each single species.
However, this is not 
the case for narrow pores where 
$L \simeq \lambda_D$ or $L \simeq a$. 
For example a positively charged 
narrow pore will typically contain more
negative ions than positive ions. The number of ions inside the channel is a fluctuating quantity in time~\cite{marbach2021intrinsic,gravelle2019adsorption}.
Surface hydrophobicity also plays a role in determining the appropriate average water density. 
In the case of highly hydrophobic patches,
even the mere presence of the 
electrolyte solution inside the pore
is questionable since vapor bubbles may 
form~\cite{beckstein2001hydrophobic,giacomello2020bubble,tinti2017intrusion}.
In MD simulations, however, the total number of atoms of each species is 
generally kept constant throughout the simulation. Hence specific strategies are required to 
ensure that the number of atoms chosen is compatible with the corresponding real system.

The standard way to tackle this issue in nanopore systems is 
to explicitly simulate reservoirs on the two sides of the pore.
This approach allows one to directly use standard tools present in common MD
packages~\cite{phillips2005scalable,hess2008gromacs,plimpton1995fast}
(\textit{e.g.} flexible cell barostats) that,
when employed during the equilibration, adapt the 
box size independently in the three directions.
Details of this method can be found, among others, in
\cite{aksimentiev2005imaging,bonome2015multistep,comer2011modeling}.
This solution requires to dedicate 
a relevant part of the computational resources to the modeling
of the reservoirs.  It is therefore useful to find alternative approaches.  

To avoid reservoirs one approach is to impose periodicity along the pore axis. This is especially suited for long
pores, where entrance effects are not dominant, and reservoirs
have a limited impact on ionic and electroosmotic flow.
This can be done, for instance, to investigate flows
in a long nanotube~\cite{qiao2003atypical}
or in a planar channel~\cite{qiao2003ion,freund2002electro} and to disentangle entrance effects. 
Thus, this approach is not applicable to biopores that are typically short and in which entrance effects are crucial, especially due to asymmetric pore entrances.
When employing this strategy,
care should be taken to select the number of atoms
constituting the confined electrolyte solution.
This is important for example to avoid
bubble formation. In planar geometries this is easily circumvented
by either imposing a mechanical 
pressure on the walls~\cite{marchio2019wetting,shahmardi2021effects},
using a barostat along the direction normal to the walls~\cite{huang2008water,chinappi2010intrinsic} or
anchoring the wall atoms to a spring and calibrating 
the constraint position to get the 
prescribed pressure~\cite{gentili2014pressure}. 
The extension of these methods to cylindrical geometries
is not straightforward and, to the best of
our knowledge, never reported in the literature.
Careful choice of the number of confined atoms is also essential to ensure that the local chemical potential is in equilibrium with that of the (non-simulated) reservoirs.
For example Widom's insertion method~\cite{widom1963some}
can be used to estimate the chemical potential and adapt the number of molecules in nanochannels~\cite{sega2015importance}, though, to the best of our knowledge, such an approach has not yet been conducted in cylindrical geometries. 

A few other, quite novel, approaches exist to remove reservoirs entirely yet mimick their effect. If the nanopore is sufficiently long\footnote{Such that the translocation time of an ion is much longer than the typical insertion time.}, the nanopore alone can be simulated~\cite{kavokine2019ionic}. 
Typically, this requires to simulate the system with several values of the number of ions $N$ and to average transport properties for different $N$ with the grand canonical probabilities of having $N$ ions in the channel (related to the free energy). In contrast if the nanopore is short, fluctuating particle numbers have to be dynamically resolved. Inserting particles in the channel with effective rates depending on the geometry of the system does not reproduce correct statistical properties of the system in general~\cite{bezrukov2000particle,marbach2021intrinsic}. Alternatively, simulation of two ``line-like domains'' (one facing the pore and one for the rest of the reservoir) yields satisfactory results~\cite{marbach2021intrinsic}.
Such strategies are quite novel, and require extensive care to be manipulated in different systems. However, they all represent promising routes to avoid explicit simulation of reservoirs.

In any case, proper account of the number of molecules within the pore is a crucial step. A possible consequence of a larger or lower number of solvent molecules or ions in the pore region, is the unrealistic 
representation of liquid layering at the 
wall. 
In particular, the effective thickness of the equivalent vacuum
layer induced by the presence of the liquid-wall interface, referred to as
the depletion layer~\cite{janecek2007interfacial}, is directly related to liquid slippage at the wall~\cite{sendner2009interfacial,gentili2014pressure},
that, in turn, can strongly affect the EOF~\cite{joly2006liquid}. Another obvious consequence is the incorrect account of electrostatic interactions and thus of the accumulated charge.  
\\

{\bf iii) Accounting for heat fluxes in non-equilibrium simulations.}
The most common approach to infer transport properties
is to perform non-equilibrium MD (NEMD) runs
where an external forcing induces the flow of the 
solution~\cite{aksimentiev2005imaging,yoshida2017osmotic,bonome2017electroosmotic,ramirez2020studying}. 
The solid walls are constrained to the simulation box and do not move.
The external forcing 
(external electric field in the case of EOF)
results in a net work on the system.
In experimental designs, systems are neither periodic nor
isolated but they are in contact with a heat bath. Hence,
the work done by the external forcing is converted 
into heat by viscous friction and eventually 
a heat flux from the system to the heat bath sets in.
MD is typically run in periodic systems with no external
boundaries (\textit{e.g.} triperiodic systems). 
To reproduce heat flows in NEMD, the common strategy is to adapt the system's equilibrium temperature control tools.
These tools, indicated as thermostats, alter the dynamics: typically in $NVT$ simulations the total energy is not a conserved quantity but the average kinetic energy of the atoms is constrained.
For a comprehensive introduction on thermostats see, among others~\cite{frenkel2001understanding,allen1989computer}.
When no net motion is present (as in equilibrium simulations)
constraining the kinetic energy amounts to
prescribing the temperature $T$.
However, in NEMD,
when a flow sets in,
the kinetic energy of the atoms is not 
equally distributed among the three 
translational degrees of freedom; it is larger
in the direction of the flow.
As a consequence, the standard usage of thermostats on all the degrees of freedom
 may artificially modify ion and solvent flows.

Several solutions have been presented to
limit possible artifacts.
One route is to couple only the solid 
atoms to the thermostat~\cite{bhattacharya2011rectification,di2021geometrically}.
Another possibility is to apply the thermostat only
to the translational degrees of freedom orthogonal to 
the flow direction~\cite{thompson2003nonequilibrium}. Note that this is possible only when studying periodic pores
where no EOF funneling at the pore entrance is present.
Finally, in some cases, the kinetic energy associated with the
streaming velocity is a very small fraction of the 
total kinetic energy~\cite{qiao2003ion,freund2002electro}. 
Hence, the application 
of a standard thermostat that acts on all the three translational 
degrees of freedom is expected to be satisfactory to  infer transport coefficients. 

To infer transport coefficients, another possibility is to simulate equilibrium systems and use linear response theory. This so-called Green-Kubo approach has been employed extensively to study flows at the nanoscale~\cite{chinappi2006molecular,yoshida2014molecular,yoshida2017osmotic}. The main advantage is that the system may be simulated at equilibrium and hence heat fluxes are entirely avoided. One common drawback however is that the method suffers from the so-called plateau problem, namely that the method requires infinitely long simulation times to converge, a problem which has only recently been solved~\cite{espanol2019solution}. Note that both NEMD and Green-Kubo approaches are equivalent in the linear regime~\cite{yoshida2017osmotic}. 
\\

{\bf Other aspects.} 
Performing MD simulations of EOF in confined geometries encompasses other challenges beyond the three discussed previously.
One additional relevant issue is the choice of the water and ion models.
Indeed, 3 points water models
do not quantitatively reproduce 
the viscosity and the diffusion coefficient of water.
For quantitatively accurate predictions,
one should rely on 4 points water models such 
as TIP4P/2005~\cite{gonzalez2010shear,abascal2005general}.
However, standard force fields for biomolecules and ions
are calibrated using 3-points water 
models~\cite{dopke2020transferability}.
Consequently there is an inevitable balance between better 
acknowledgment of biomolecule-water interactions
or water viscosity and diffusion.
A similar problem occurs for ions.
Standard rules for non-bonded interactions 
(Lorentz-Berthelot~\cite{allen1989computer})
do not reproduce quantitatively correct ionic
conductivities and {\sl ad hoc} corrections to force fields
have been recently proposed
(\textit{e.g.} CUFIX~\cite{yoo2011improved} correction for
CHARMM~\cite{huang2017charmm36m}).

Another debated topic is how to apply the 
external forcing.
In typical experimental conditions,
a voltage drop $\Delta V$ is applied between two 
reservoirs that are separated
by the nanoporous membrane.
The distance from the membrane to the electrode
is typically orders of magnitude larger than
the pore diameter and, hence,
as a first approximation, the 
reservoirs can be considered infinite.
On the other hand, fast computation 
of electrostatic interactions in 
MD simulations requires periodic boundary conditions~\cite{essmann1995smooth}
so, in essence, any simulated system is not 
a single pore but a series of arrays of pores. This is obviously incompatible with fixed voltage boundary conditions.
The usual solution 
is to apply a constant and homogeneous 
electric field ($E$) parallel to the pore axis
and wait for the migration of ions towards the membrane.
It has been shown that this solution is equivalent
to the application of a voltage drop $\Delta V = L_z E$ 
where $L_z$ is the size of the periodic box in the 
axial direction~\cite{gumbart2012constant}. 

\begin{figure}
	\center
	\includegraphics[width=\linewidth]{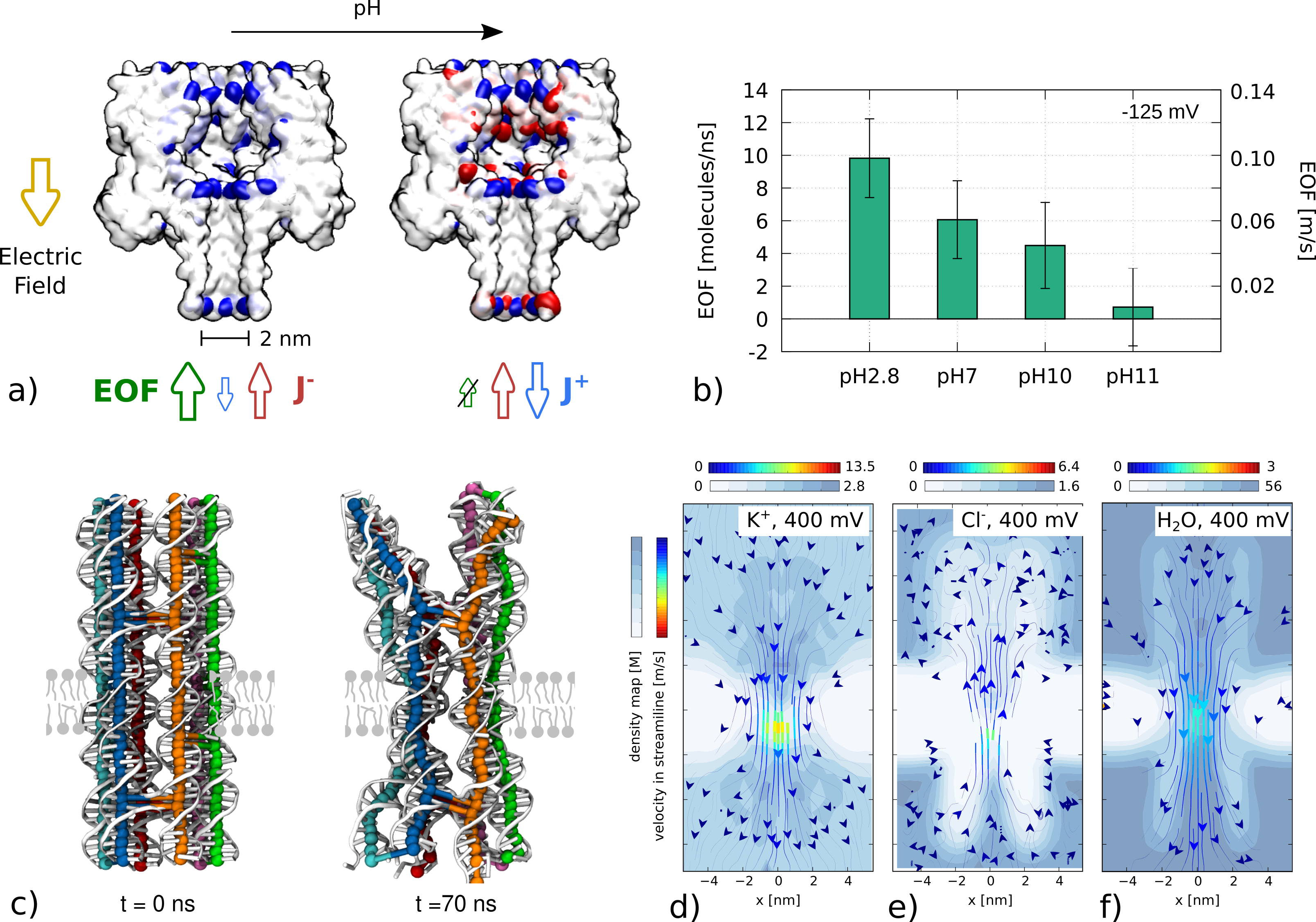}
	\caption{
		{\bf Biological Nanopores.}
		{\bf a) $\alpha$-Hemolysin ($\alpha$HL) at different pH.}
		Titratable residues 
		in the interior surface affect the pore 
		selectivity~\cite{bonome2017electroosmotic,betermier2020single}.
		The positively and negatively charged residues
		are in blue and red, respectively.
		At low pH (left), the acididic residues are almost all protonated (neutral), 
		and the pore interior is mainly positively charged.
		The pore is thus anion selective and an EOF sets in,
		in the direction opposite to the electric field.
		At high pH (right), acidic residues and 
		tyrosines (pKa~$\sim$~10.5) become negative,
		and some basic groups become neutral (histidines, N-terminals).
		Consequently, the overall $\alpha$HL interior becomes neutral.
		and EOF vanishes.
		{\bf b)} EOF from MD simulations for
		$\alpha$HL in 1M NaCl water solution 
		for $\Delta V=-125\,$mV at different pH. 
		Data for pH=7 and pH=10 are taken from~\cite{betermier2020single};
		data for pH=2.8 and pH=11 are original data 
		obtained by using the same protocol,
		(protonation states predicted by H++ server~\cite{anandakrishnan2012h++}).
		{\bf c) DNA-origami nanopore.}
		Cartoon representation (gray) of the initial (left) and equilibrated (right) DNA structure,
		overlaid with a chickenwire representation (colors). 
		In the chickenwire representation, 
		beads indicate the locations of the centers of mass of individual basepairs; 
		horizontal connections between pairs of beads indicate interhelical crossovers.
		The lipid surrounding the DNA is sketched in grey,
		and the 1M KCl water solution atoms filling the simulation box are not shown.
		{\bf d-f)} Local density (gray scale) and local velocity (streamlines) of 
		$\mathrm{K}^+$ ({\bf d}) and
		$\mathrm{Cl}^-$ ({\bf e}) ions and
		water ({\bf f}) through the system of panel {\bf c} at 400 mV.
		Panels {\bf c-f} are adapted from Yoo and Aksimentiev~\cite{yoo2015molecular}.
		}
	\label{fig:biological}
\end{figure}

\subsection{Applications to study EOF in biological nanopores}
Here we report two examples of MD simulations of EOF in biological pores.

{\bf$\alpha$-Hemolysin.}
The most widely studied biological nanopore 
through MD 
is $\alpha$-Hemolysin ($\alpha$HL),
whose first simulation was reported by
Aksimentiev and Schulten~\cite{aksimentiev2005imaging}.
The pore was characterized in terms of permeability for water and ions
at standard temperature and pressure.
The system was composed by $\sim 300\,000$ atoms,
including the protein, the lipid bilayer and a 1M KCl water solution.
By applying an external electric field, 
an ionic current was established.
The measured total current and ionic selectivity were 
in excellent agreement with available experimental data,
demonstrating the capability of non-equilibrium MD simulations to 
predict with quantitative accuracy ionic currents through
transmembrane biological pores induced by applied voltages.
Afterwards,
many studies were published using a similar all-atom setup
to obtain molecular insight on current 
blockage due to macromolecules inside the pore 
(DNA, proteins)~\cite{wells2007exploring,
mathe2005orientation,martin2009determination,di2019insights}
and to study the effect of different salt types,
concentration,~\cite{bhattacharya2011rectification}
or protonation state of the 
exposed residues~\cite{bonome2017electroosmotic,betermier2020single}
on ionic selectivity and EOF.
At neutral pH, 
several charged residues (both positive and negative) 
are present in the pore interior
and $\alpha$HL is slightly anion selective.
The charge of such residues can be altered by varying the 
pH of the solution, see Fig.~\ref{fig:biological}a.
In particular, 
at low pH the pore interior is mostly positively charged, 
as aspartic acid and glutamic acid are almost all neutralized, 
whereas histidine, lysine and arginine have a positive charge.
In such conditions, 
anions accumulate in the pore resulting in 
a strong unbalance of positive and negative ionic fluxes
and an intense EOF,
see Fig.~\ref{fig:biological}b.
Conversely, at high pH, acidic residues and tyrosines
(hydroxyl group, pKa~$\sim$~10.5) become negatively charged,
while some basic groups, such as histidines and N-terminals 
(pKa~$\sim$~8.7) become neutral. 
The reduction of the net charges of the pore surface
leads to a reduction of the pore selectivity and, consequently, 
the EOF is negligible, Fig.~\ref{fig:biological}b.
This scenario is in fair agreement with experiments 
by Asandei et al.~\cite{asandei2016electroosmotic}, 
where a positively charged peptide was captured 
in $\alpha$HL against electrophoresis at pH=2.8,
while it is not captured at pH=7, in agreement with EOF strengthening at low pH.

{\bf DNA-origami nanopore.} 
Another interesting class of emerging {\sl artificial} biopores
are made of DNA-origami 
nanostructures~\cite{li2015ionic,yoo2015molecular,langecker2012synthetic,burns2013self}.
When decorated with hydrophobic anchors, 
self-assembled DNA structures can 
spontaneously merge with a lipid bilayer membrane, 
forming a transmembrane nanopore,
see Fig.~\ref{fig:biological}c. 
DNA-origami 3D structures can be rapidly designed {\sl in silico}
using, for instance, the caDNAno software~\cite{douglas2009rapid}, 
and further modeled and modified through 
standard biomolecular tools~\cite{yoo2015molecular}.
In comparison to conventional nanofabrication approaches, 
DNA self-assembly offers an efficient way to control nanopores with
subnanometer resolution and massive parallelization.
Furthermore, 
many chemical modifications can be
incorporated within the DNA membrane channels,
when compared to the more limited options of other assembly systems composed
of peptides or proteins~\cite{burns2013self,burns2013lipid,howorka2011rationally}.
Since DNA is highly negatively charged, 
the pore is expected to be cation selective.
MD simulations confirm this scenario
(both fluxes and concentrations inside the pore are larger for $\mathrm{K^+}$ with respect to $\mathrm{Cl^-}$, Fig.~\ref{fig:biological}d-e) and reveal an intense EOF.
For 400 mV, 
the authors report a water flow of about 120 molecules/ns~\cite{yoo2015molecular}
corresponding to an average electroosmotic velocity 
of $v_{eo}=0.286\,$m/s at the center of the channel.
Assuming a linear dependence of $v_{eo}$ with the applied voltage, 
it can be noted that the resulting EOF is comparable to the low pH $\alpha$HL
(DNA-origami $v_{eo} \sim 0.09$~m/s for 125 mV)
see Fig.~\ref{fig:biological}d-f.
In the same work~\cite{yoo2015molecular}, it has also been found
that the conductance of DNA channels depends on membrane tension, 
making them potentially suitable for force-sensing applications.

%% file: mesoscale.tex
\section{Mesoscale methods}\label{sec:meso}

As discussed in Section~\ref{sec:md}, Molecular Dynamics simulations encompass numerous effects relevant to nanoscale electroosmosis modeling: thermal fluctuations, detailed interactions of the fluid with the confining walls, hydrodynamic interactions between translocating particles and with the walls and electrostatics including the effect of the dielectric medium. 
The effect of confinement arises naturally in all-atoms approaches and depends only on the forces between atoms, while it has to be specifically implemented in continuum methods, whose coarse-grained parameters (\textit{e.g.} viscosity, dielectric constant, electrical conductivity) are typically available as constants representing bulk values. In addition, continuum models as discussed in Section~\ref{sec:continuum}, for instance do not include thermal fluctuations, which are critical in nanopores. 

\begin{figure}
	\center
	\includegraphics[width=\linewidth]{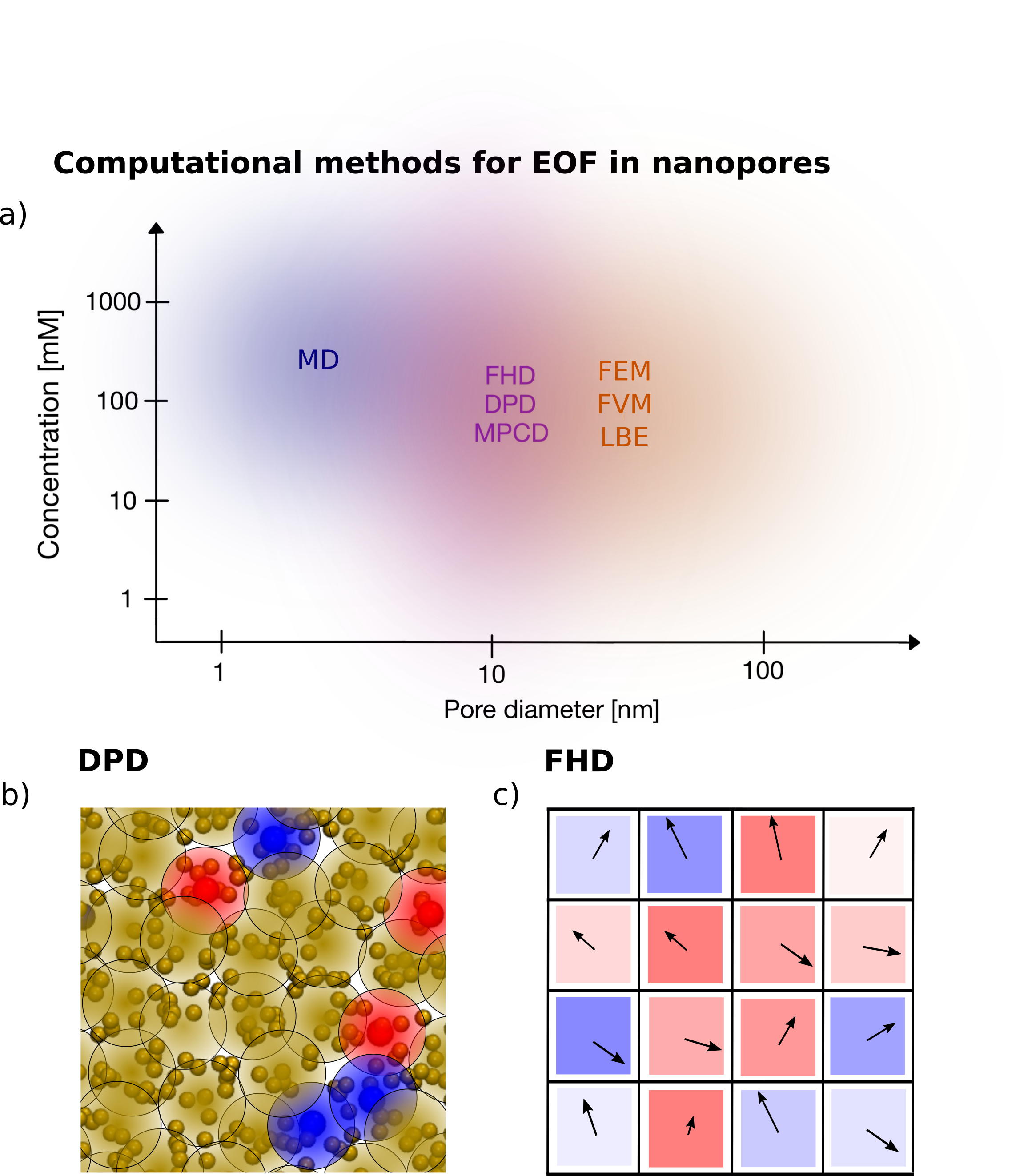}
	\caption{
	\textbf{Methods to simulate EOF. a)} Diagram of the different computational methods available to simulate EOF in nanopores, classified according to the pore size and concentration of the simulated system. 
	The colored area surrounding the name of the method shows the range of systems which are most suited to be simulated.
	The areas are shaded since the choice has to take into account additional system-specific factors, \textit{e.g.} presence of moving particles, importance of thermal fluctuations, importance of ion-specific effects. 
	The blue area represents an atomistic method, all-atom MD. 
	The violet area represents a group of mesoscale methods which do not have atomistic resolution but still are able to model the thermal fluctuations of the system, namely Fluctuating Hydrodynamics, Dissipative Particle Dynamics and Multi-Particle Collision Dynamics.
	The orange area refers to different computational methods to solve PNP-NS equations, such as Finite Volume Method, Finite Element Method and to Lattice Boltzmann Electrokinetics.
	\textbf{b)} Sketch of the simulation set up of a bottom-up approach (DPD). 
	Each DPD particle is represented by a big blob delimited by a black circle, which can be yellow (neutral solvent), blue (positively charged) or red (negatively charged). 
	The blobs overlap as a consequence of the weak repulsive interactions typical of DPD simulations.
	In the background, a possible atomistic system represented by the DPD particles, with solvent atoms (yellow), positive ions (blue) and negative ions (red).
	\textbf{c)} Sketch of the simulation set up of a top-down approach (FHD). The fluid is divided into cells, each of which has a fluctuating velocity, here represented as an arrow, and a charge arising from the ion density fluctuations, here represented as a colored box, for positive (blue) and negative (red) charges.
	}
	\label{fig:meso}
\end{figure}

The downside is that atomistic simulations are limited in the spatial and temporal scales that can be resolved. Even with the ever increasing availability of computational resources an intermediate scale exists (\textit{i.e.} mesoscale) that requires more details than continuum methods to be described and yet is out of reach of atomistic approaches. 
Any technique suitable to tackle such problems falls in the quite broad category of {\sl mesoscale methods}. 
Mesoscale methods represent a heterogeneous set of techniques, that sometimes have little in common among one another. For a given problem, a specific method is usually better suited than another, depending on factors such as the geometry of the system, the boundary conditions and the presence of moving nanoparticles or biomolecules. 
In the following paragraph, we focus on mesoscale techniques that can be used to simulate nanoconfined electrolyte solutions and electroosmotic flows, summarized in Fig.~\ref{fig:meso}a. 

The diversity of mesoscale approaches calls for additional sub-classifications. One common discriminating feature is the model used to describe the fluid.
This can be done as an extension of continuum methods, describing the fluid in terms of a velocity field, where properties such as viscosity and dielectric constant can be directly assigned. To account for finite-size effects, modifications to the dynamical equations or to the local properties can be added. We refer to these approaches as {\sl top-down} models. Another possibility is to extend on molecular dynamics, and model the fluid by a set of mobile, coarse-grained particles that represent molecules or groups of molecules rather than single atoms. Interactions between fluid particles result in effective fluid properties, that are naturally modified by the confinement. We refer to these approaches as {\sl bottom-up}.
In the following, we focus on representative methods belonging to either category
(Dissipative Particle Dynamics (DPD) for {\sl bottom-up} and Fluctuating Hydrodynamics (FHD) for {\sl top-down}) 
and provide a brief overview of some of other approaches. 
More insight, especially on these other mesoscale approaches, 
may be found for example in~\cite{rotenberg2013electrokinetics}.

%

\subsection{Bottom-up approaches}

Bottom-up approaches model the fluid as a set of interacting particles, whose position and velocities are updated according to rules which preserve total momentum, see Fig.~\ref{fig:meso}b.
The hydrodynamic behavior of the fluid hence arises from the microscopic interactions between the mesoscale particles.

{\bf Dissipative Particle Dynamics.}
The DPD model was introduced by Hoogerbrugge and Koleman in 1992~\cite{hoogerbrugge1992simulating} as a stochastic differential equations model. Its statistical mechanics properties were studied a few years later by Espa\~nol and Warren~\cite{espanol1995statistical}, 
who obtained a fluctuation-dissipation relation between the deterministic and stochastic parts of the equations. 
The original DPD set of equations determining the motion of all the particles representing the fluid reads
\begin{align}
\frac{d\boldsymbol{{x}}_i}{dt}=\boldsymbol{v}_i\;,\\
\label{eq:DPD}
m\frac{d\boldsymbol{{v}}_i}{dt}=\sum\limits_{j\ne i}\boldsymbol{F}^C_{ij}+\boldsymbol{F}^D_{ij}+\boldsymbol{F}^S_{ij}\;,
\end{align}
where $\boldsymbol{x}_i$ is the position of the $i^{\mathrm{th}}$ particle,
$\boldsymbol{v}_i$ its velocity, 
$m$ its mass and $\boldsymbol{F}^C_{ij}$, $\boldsymbol{F}^D_{ij}$, $\boldsymbol{F}^S_{ij}$ are 3 pairwise radial forces, respectively a weakly repulsive conservative force, a dissipative force proportional to the radial component of the velocity difference between particles, and a stochastic force proportional to a white noise process~\cite{espanol2017perspective}. 
The magnitude of the three forces is determined by parameters which define the properties of the fluid, and decreases with the distance between the particles, vanishing if the interparticle distance is larger than a cutoff radius.

The formulation of DPD has been modified in several ways to represent increasingly more complex fluids and systems.
First, in the original model, the conservative force decreases linearly with the interparticle distance, limiting the range of equations of state of the modeled fluid.
Modifications to the conservative force introducing a local density variable allow fluids with different equations of state to be simulated~\cite{pagonabarraga2001dissipative,warren2001hydrodynamic,trofimov2002thermodynamic}.
Second, the addition of an internal energy variable for each particle and a stochastic equation governing energy exchange between particles enables simulations with temperature gradients and heat flows~\cite{avalos1997dissipative,espanol1997dissipative,tong2020study}.
Finally, interactions with fixed DPD particles representing the solid surface enable simulations of nanoscale flows, including hydrodynamic interactions between colloids and the effect of confining walls with different slip lengths~\cite{boek1997simulating,keaveny2005comparative,tiwari2008simulations,li2008hydrodynamic,filipovic2011dissipative,gubbiotti2019confinement}.

Two different strategies can be adopted to represent an electrolyte solution and study charged systems with DPD. One possibility is to use charged DPD particles, directly representing the ions. Dynamics are given by the original DPD equations, with the addition of Coulombic forces in Eq.~\eqref{eq:DPD}. Simulations of electrolyte solutions including polyelectrolytes have been conducted~\cite{lisal2016self}, even in an electroosmotic planar channel setting~\cite{smiatek2011mesoscopic}. Smoothed charges are used to avoid artificial ion pairing due to the weak repulsive interactions $\boldsymbol{F}^C_{ij}$ compared to the Coulombic attraction~\cite{warren2014screening}. Polarizability effects can also been included, employing coarse-grained DPD water molecules or proteins~\cite{peter2014polarizable,peter2015polarizable}.

Another possibility to model electrolyte solutions is to consider the DPD particle as a portion of solution rather than as a group of solvent molecules or a single ion.
This requires to represent the quantity of ions carried by the DPD particle as an independent variable with additional evolution equations~\cite{deng2016cdpd,ehdpd}, an approach which has proven useful for the modeling of solutions in advection-diffusion-reaction problems~\cite{li2015transport} and can also provide control on the fluid equation of state including ion-specific effects such as excluded volume~\cite{ehdpd}.
Using a separate evolution equation for the ion dynamics has an advantage in terms of scalability of the system size since multiple ions can be represented by a single DPD particle. Ion transport models are also more versatile as they are governed by an independent set of equations and parameters.

While noteworthy progress has been made in recent years to simulate electrolyte solutions in confined environments with DPD, some issues remain to be tackled.
Most importantly, tuning transport coefficients, the fluid's equation of state, slippage and polarizability is crucial to accurately model the fluid's properties. Yet, it is not straightforward due to the bottom-up nature of DPD: forces either have to be derived from atomistic models via a coarse-graining procedure~\cite{han2021constructing} or calibrated to map fluid properties~\cite{groot1997dissipative,boromand2015viscosity,gubbiotti2019confinement,ehdpd}. Reproducing the properties of existing fluids is critical to widen the set of fluids accessible with DPD. This requires additional efforts to improve both DPD models of electrolyte solutions and the calibration of the fluid properties.

{\bf Multi-Particle Collision Dynamics.}
Another possible bottom-up approach is Multi-Particle Collision Dynamics (MPCD)~\cite{gompper2009multi}. 
Here the fluid is represented as a set of particles with position and velocity free to move continuously in space. 
Particle velocities are updated via a streaming step in which the positions are updated and a collision step in which the fluid domain is divided into cells of constant size, and 
the velocities of all the particles in each cell are rotated by a random vector which is different for each cell, but identical for all particles in the same cell.
Solute particles may be embedded into the MPCD description by treating them as fluid particles interacting with each other, propagating 
their positions according to standard MD algorithms instead of having a simple streaming step~\cite{ceratti2015stochastic}.
In this way, effects due to the finite size of ions can be taken into account.  
MPCD has been used to study nanoscale systems such as polymer translocation dynamics through nanopores~\cite{katkar2018role} and EOFs in planar channels~\cite{ceratti2015stochastic}.

{\bf Lattice-Boltzmann.}
In the Lattice-Boltzmann (LB) approach the fluid is modeled as a set of particles with position and velocity. One key difference here with respect to other approaches is that the particles evolve on a lattice and represent the probability density function (or density) of the fluid ~\cite{aidun2010lattice}.
Similarly to MPCD, the dynamics that determine the fluid behavior include (\textit{i}) a collision step in which particle velocities are updated according to a collision operator that preserves total momentum, and (\textit{ii}) a streaming step in which the distribution of particles evolves according to their velocities.
The LB method has been applied to simulate electrolyte solutions at the 
nanoscale~\cite{melchionna2004electrorheology,rotenberg2013electrokinetics} and electrokinetic phenomena, combining LB for hydrodynamics and the resolution of convection-diffusion equations for charges, an approach known as Lattice Boltzmann Electrokinetics (LBE)~\cite{capuani2004discrete,obliger2013numerical,asta2019lattice}.

\subsection{Top-down approaches}
Shifting perspectives, top down approaches describe fluid flow in a continuous fashion, see Fig.~\ref{fig:meso}c, with enhanced fundamental equations to include details that are relevant at the nanoscale.
An important tool to account for fluctuations at the nanoscale is called Fluctuating Hydrodynamics (FHD). We review how to use this tool in the context of electroosmosis below, and give some brief insight into other top down approaches.

{\bf Fluctuating ionic concentration field in fluctuating hydrodynamic field.}
The first step is to introduce fluctuations in the hydrodynamic equations for conservation of momentum and mass balance Eqs.~(\ref{eq:momentum}) and (\ref{eq:incompressible}), 
as in~\cite{de2006hydrodynamic,peraud2016low}
\begin{align}
\label{eq:FHD}
	&\rho\frac{\partial\boldsymbol{u}}{\partial{t}}+\rho\boldsymbol{u}\cdot\boldsymbol{\nabla}\boldsymbol{u}=\boldsymbol{\nabla}\cdot\boldsymbol{\Sigma} + \boldsymbol{\nabla} \cdot \bm{S} -\rho_e\boldsymbol{\nabla}\Phi\ \;,\\
    & \boldsymbol{\nabla}\cdot\boldsymbol{u}=0
\end{align}
where $\boldsymbol{\Sigma} =-p\boldsymbol{I}+2\eta\boldsymbol{\sigma}$ is the stress tensor defined in Eq.~\eqref{eq:newt}, $S$ is a random stress tensor satisfying the fluctuation-dissipation theorem, whose components are white in space and time, with mean zero, and $\langle S_{ij}(\bm{r},t)S_{kl}(\bm{r}',t')\rangle=2\eta k_B T (\delta_{ij}\delta_{kl} + \delta_{il}\delta_{jk})\delta(\bm{r}-\bm{r}')\delta(t-t')$. Note that here the fluid density is assumed to be uniform and constant but extensions are possible~\cite{peraud2016low}. 
The advantage of Eq.~\eqref{eq:FHD} is to keep a continuous, high-level description of the fluid, that does not rely on resolving individual molecular collisions, or on maintaining a constant temperature heat bath. Yet, Eq.~\eqref{eq:FHD} does describe thermal fluctuations at small scales, conserves momentum, and is consistent with fluctuation-dissipation.  Note that we keep the non-linear term in the Navier-Stokes equation for now; as in non-equilibrium cases typical of electroosmosis, equations have to be linearized around non-zero average velocity fields. 

Numerous studies show an important impact of fluctuations in confinement as modeled through Eq.~\eqref{eq:FHD}.  The equations may be solved analytically in specific cases to investigate overall transport quantities. For example, for confined fluids, fluctuations impact the de-wetting transition~\cite{mecke2005thermal} and bubble formation by cavitation~\cite{gallo2018fluctuating}. Memory effects in the center of mass diffusion have also been identified~\cite{detcheverry2012thermal}. Finally fluctuations of the surrounding confining material can enhance ensemble diffusion~\cite{marbach2018transport}. 
More generally, the equations are amenable to numerical solutions of the fluid velocity field on a grid. Such numerical solutions require careful discretization of the noise operator~\cite{balboa2012staggered}, in particular to properly account for confining boundaries~\cite{sprinkle2017large}, and of the forces at play, especially when looking at fluid-structure interactions where the structure also has thermal fluctuations~\cite{atzberger2007stochastic,atzberger2011stochastic}. 

To describe ions in such fluctuating fields, at low enough densities, one can use continuous concentration fields. As for the fluid, we can update the continuity equation Eq.~\eqref{eq:continuity} for each species $\alpha$ with a stochastic term~\cite{de2006hydrodynamic,donev2019fluctuating,pagonabarraga1997fluctuating}
\begin{equation}
    \frac{\partial c_\alpha}{\partial t} = - \bm{\nabla}\cdot ( c_\alpha \bm{u} + \bm{j}_{\alpha} + \bm{F}^S_\alpha)
\end{equation}
where we take $\bm{j}_{\alpha}$ to be given by the constitutive ion flux Eq.~\eqref{eq:nernstplanck} and $\bm{F}^S_{\alpha}$ a stochastic flux that satisfies fluctuation-dissipation, 
\begin{equation}
    \bm{F}^S_{\alpha} = \sqrt{2 D_\alpha c_\alpha} \bm{\mathcal{Z}}
\end{equation}
where $\bm{\mathcal{Z}}$ is a Gaussian vector with uncorrelated components in space and time. 
The fluctuating correction can also be recovered through density functional theory~\cite{dean1996langevin}.

Such a formulation is interesting as it allows one to efficiently investigate fluctuation-driven behavior, both numerically and analytically. 
This approach has been used to identify enhanced charge transport due to fluid fluctuations, that result in ion concentration corrections to diffusion, sometimes leading to negative diffusion~\cite{donev2019fluctuating} or enhanced electric conductivity~\cite{peraud2017fluctuation}. In other contexts, giant fluctuations may appear at interfaces with concentration gradients, that can be enhanced by applied electric fields~\cite{peraud2016low}.

The effect of fluctuations on electroosmosis has not yet been assessed in detail, and numerous questions are open. 
Such problems are now accessible to simulations, as confining boundary conditions have already been successfully implemented~\cite{donev2019fluctuating2} and EOFs have been simulated in FHD for ionic liquids~\cite{klymko2020low}. One might expect fluctuation enhanced behavior of the EOF, as identified for the electrical conductivity~\cite{peraud2017fluctuation}, yet the mechanisms remain to be unraveled. Furthermore, FHD represents a good starting point to rationalize further the effect of fluctuations at the boundaries, for example originating from charge regulation on surfaces~\cite{secchi2016scaling,kim2005electroosmosis}, which should be crucial for EOF. Overall this is a completely open question where combined mathematical frameworks and numerical investigations have to be built. 

{\bf Particle ionic field in fluctuating hydrodynamic field.}
At high ionic densities, or in strongly confined systems, continuum approaches are not physically or numerically relevant. In fact, since the Debye length scales as $\lambda_D \sim c^{-1/2}$, where $c$ is the ionic concentration, the number of ions in a grid cell is at most $\lambda_D^3 c \sim c^{-1/2}$ (since the discretization grid length is at least smaller than $\lambda_D$). Hence, numerically, the number of ions per cell at high concentrations is too small for a continuum description of ions to be sensible. Notice, that, physically, at very high concentrations, $\lambda_D$ approaches the molecular scale $\lambda_D \lesssim 1~\mathrm{nm}$ and a continuum description with cell sizes smaller than $\lambda_D$ is also questionable at that scale.
Therefore, at high concentrations, a natural top-down simulation improvement is to resort to a discrete description for the ions~\cite{ladiges2021discrete,wu2015simulation}, but keep a continuous description for the solvent velocity field. 
To avoid common difficulties associated with tracking a particle of finite size in fluid flow 
most works rely on methods such as Immersed Boundaries (IB)~\cite{mittal2005immersed,delong2014brownian,fadlun2000combined} or fluctuating force coupling methods~\cite{lomholt2003force,delmotte2015simulating}, which effectively smear the particle on the velocity grid. This forms a hybrid, top-down approach. 
While such mixed approaches remain quite novel, especially in their applications to charged species~\cite{ladiges2021discrete}, they could capture effects that occur on the molecular scale due to fluctuating hydrodynamics, especially as they are well suited to bridge the scales between large reservoirs and confined pores.

{\bf Smoothed Dissipative Particle Dynamics.} Another top-down approach is a top-down version of DPD, known as Smoothed Dissipative Particle Dynamics (SDPD)~\cite{espanol2003smoothed}. It consists in a discretization of the Navier-Stokes equations into a particle system instead of a fixed mesh, in which each particle moves according to its velocity, while the continuous velocity field (and any other relevant field) can be interpolated using bell-shaped functions centered on each particle.
The momentum balance equation can then be used to update the velocity of each particle.
This approach can be considered as a Lagrangian discretization of the Navier-Stokes equations, that includes thermal fluctuations in a consistent way. The advantage with respect to DPD techniques is that properties such as viscosity are inputs of the model and don't have to be calibrated. The introduction of charged species in the model would require either considering some charged fluid particles as ions (hence in a mixed bottom-up/top-down approach), similar to what is done in DPD with explicit ions~\cite{smiatek2011mesoscopic}, or the introduction of an ionic concentration variable (somehow similar to DPD approaches in~\cite{deng2016cdpd,ehdpd}) with an additional set of equations based on the continuity equation Eq.~\eqref{eq:continuity}.

%% file: techno.tex
\section{Technological Applications of EOF}
\label{sec:techno}
From a technological perspective, EOF is attractive as it provides the unique possibility to induce and control a flow by easily tuning the externally applied voltage drop. 
Compared to standard pressure driven flux, this is experimentally extremely convenient at the micro to nanoscale as it avoids common issues such as membrane fracture~\cite{book_micro}. EOF can be further manipulated by modifying the system design through the surface properties of the channel walls as discussed in section~\ref{sec:3routes}.

In the last decades this has led to a large variety of applications of
EOF at the microscale such as pumps~\cite{wang2009electroosmotic}, 
micromixers~\cite{chang2007electrokinetic},
and transdermal drug delivery systems~\cite{kusama2021transdermal}.
Some of these applications have been recently extended down 
to the nanoscales, such as pumps~\cite{ai2010low,wu2016alternating,wu2018chemoresponsive}. 
Nanofluidic pumps can be used in direct current mode~\cite{ai2010low} 
exploiting intrinsic selectivity due for instance to fixed charge accumulation at the pore walls 
as  in Fig.~\ref{fig:mechanism}a-c. 
They can also be used in alternate current mode, where, 
to achieve a net EOF flow after a voltage cycle, 
current rectification is provided by 
pore asymmetry~\cite{wu2016alternating}, 
induced charge~\cite{di2021geometrically} or
unequal ions mobilities~\cite{amrei2018oscillating,bukosky2019extreme}.
The latter does not necessarily require an asymmetric porous membrane 
and, consequently, could simplify the fabrication process.
This possibility extends 
the variety of technological solutions aimed at 
harnessing EOF and remains to be explored experimentally. 
Further interesting recent
progress in EOF includes
chemoresponsive pumps where 
the flow decreases when a given 
species reaches a threshold concentration~\cite{wu2018chemoresponsive}
and pumps integrated in flexible materials such as fabric~\cite{bengtsson2017large}.
Moreover, other purely nanoscale applications 
have emerged such as 
the development of soft actuators~\cite{ko2020electroosmosis}
and the design of a nanorobot able to move along a solid-state
surface~\cite{si2020nanoparticle}. 

In the following sections, we focus on 
one specific application of EOF at the nanoscale,
namely the electrohydrodynamic manipulation of
nanoparticles and molecules to control their interaction
with a nanopore or a nanoporous membrane. 
This is particularly relevant in nanopore sensing devices, see Section~\ref{sec:sensing},
and it may also be important for applications to blue energy and desalination, see Section~\ref{sec:blue}.

\begin{figure}
        \center
        \includegraphics[width=\linewidth]{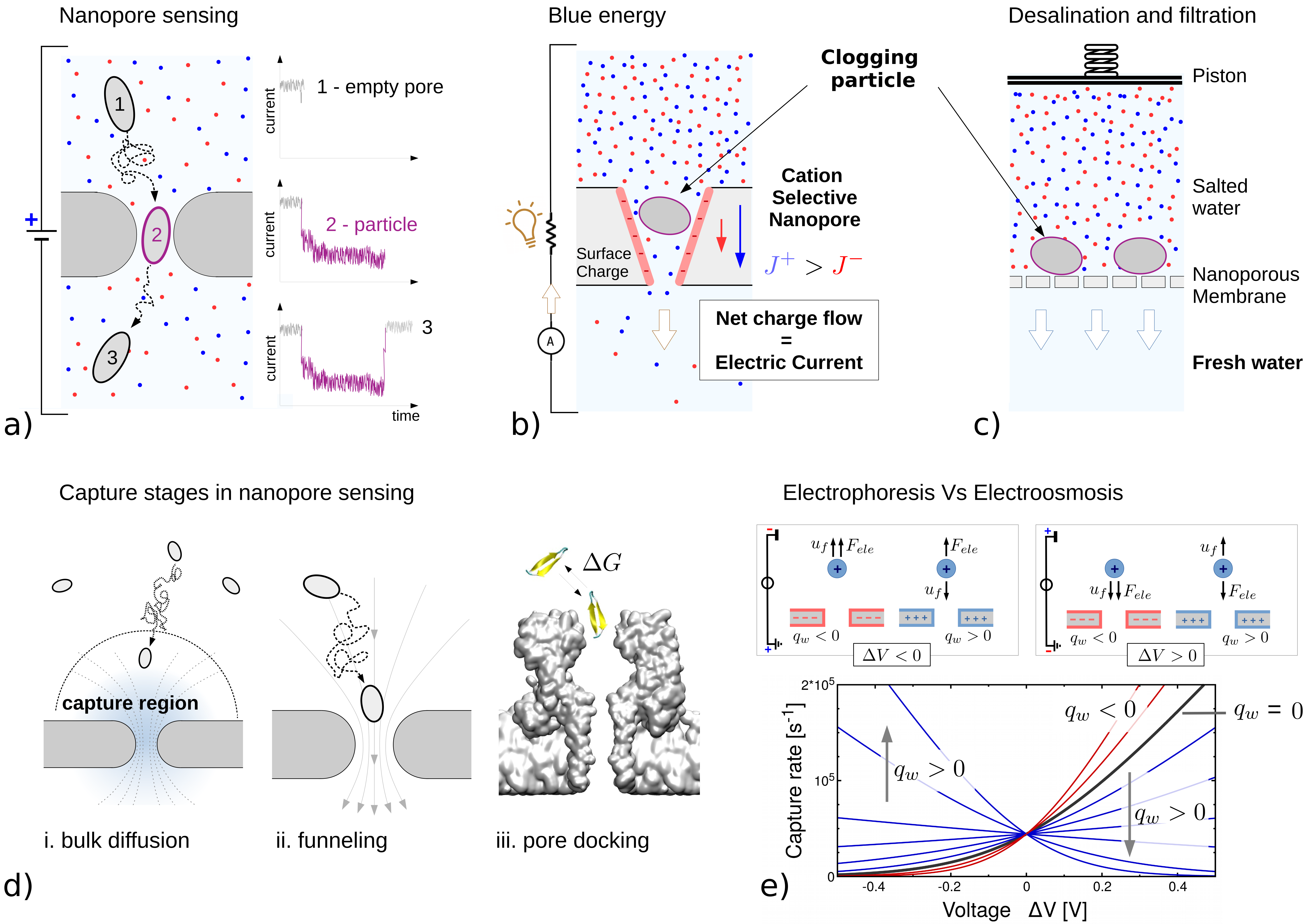}
        \caption{{\bf Nanopore/nanoparticle interactions: the role of EOF in applications.}
		{\bf a) Nanopore sensing.}
		The properties of a translocating particle are deduced
		from the signature it leaves in the electric current trace.
		Indeed, when the particle enters the pore, it partially blocks the
		ion flow and, consequently, the electric current is reduced (stage 2, violet).
		The 
		capture of molecules and particles to be analyzed needs
		to be favored. 
		This may be particularly challenging 
		when particles are not charged. In this respect, EOF
		constitutes a promising approach since it may induce
		particle capture regardless of their charge.
		{\bf b) Blue energy and c) Filtration.}
		The interaction between dispersed nanoparticles
		and nanoporous membrane may impact the 
		performance of desalination, filtration and blue energy harvesting
		membranes since nanoparticles can clog the pores ({\sl fouling}).
		EOF can be employed to generate a flow to clean the membrane
		({\sl backwashing}).
		{\bf d) Particle capture.}
		The three phases of the capture in a nanopore
		sensing device. 
		{\sl i)} Bulk diffusion.
		Far from the pore, 
		electrophoresis and electroosmosis 
		are negligible. Hence, the dynamics are dominated by Brownian motion, 
		until the particle reaches the pore capture region.
		{\sl ii)} Funneling. Once the particle is in the capture region, 
		Brownian motion competes with electric and hydrodynamic
		forces. If the latter are directed towards the pore, 
		the particle experiences a funnel-like field: 
		the closer the particle to the pore entrance, the larger the attractive force. 
		{\sl iii)} Pore docking. The particle finally reaches the pore entrance region, 
		where molecular pore-particle interactions become relevant.
		{\bf e) Capture frequency} of a positively charged particle,
		as a function of $\Delta V$ from a theoretical model~\cite{chinappi2020analytical}
 		showing the competition/cooperation between EOF and electrophoresis 
		depending on the wall surface charge $q_w$.
		The black line refers to pure electrophoresis ($q_w = 0$), 
		red and blue lines to positively and negatively charged walls.
		Panel d-e adapted from~\cite{chinappi2020analytical}.
		Panel d-iii was realized using the VMD software~\cite{humphrey1996vmd}.
                }
        \label{fig:tec}
\end{figure}

\subsection{EOF enhanced nanopore sensing}
\label{sec:sensing}

One of the most widely studied applications of EOF at nanoscale is nanopore sensing.
Nanopore sensing is based on the presence of particles 
inside the pore~\cite{varongchayakul2018sensing,bayley2000stochastic}.
The molecular properties of analytes are 
inferred by the alteration of
the electric current flowing through 
the nanopore~\cite{chinappi2018protein,cressiot2012protein,luo2014resistive} (Fig.~\ref{fig:tec}a). 
This approach is usually referred to as \textit{resistive pulse} 
and it is, by far, the most commonly employed strategy for nanopore sensing. 
Other
approaches have been proposed, such as acquisition of the transverse tunneling current along the
substrate plane~\cite{di2016decoding}, fluorescence signals 
elicited by a laser light~\cite{ohayon2019simulation} and FRET spectroscopy 
to sense the deformation of a biopore due to the molecule's passage~\cite{spitaleri2021adaptive}.
Also plasmonic nanostructures have been recently used 
in nanopore sensors enhancing optical
spectroscopy, local temperature control
and molecule thermophoresis~\cite{garoli2019plasmonic,tarun2019spatiotemporal}.

A sketch of the resistive pulse approach is 
reported in Fig.~\ref{fig:tec}a. 
When the particle is outside the pore, 
the electric current fluctuates 
around a base level (stage 1, grey). 
When the particle enters the pore, it partially blocks the 
ion flow and, consequently, the electric current is reduced (stage 2, violet).
When finally the particle leaves the pore,
the current returns to its empty pore value (stage 3, grey).
There are three main requirements that a nanopore 
sensing device needs to fulfill. 
First, molecules dispersed 
in the reservoirs must enter the nanopore 
(capture control).
Then, the captured molecule must reside inside the pore 
for enough time to record a stable 
current signal (translocation/residence control)
and, finally, 
signals coming from different molecules 
(or different parts of the same molecule, 
{\sl e.g.} single monomers when the aim is polymers sequencing~\cite{chinappi2018protein})
must be different from each other (distinguishability).
EOF has been shown to be a useful method to control the 
capture, 
as it will be described in the following.

The capture of a molecule into
a nanopore is a complex process governed by the interplay
among Brownian diffusion, hydrodynamics, 
physico-chemical and electrostatic effects. 
In general, we can roughly split the capture processes in three main phases,
see Fig.~\ref{fig:tec}d,~\cite{chinappi2020analytical}.
(i) Bulk diffusion: 
Far from the pore, 
the electroosmotic drag and electrophoresis are, in essence, negligible. 
Here, the particle dynamics are dominated by Brownian motion.
(ii) Funneling: 
Brownian motion occasionally 
brings a particle so close to the pore 
that electric and hydrodynamic forces
start to be relevant.
Supposing that the resultant of these actions 
is directed toward the pore, the
particle experiences a funnel-like field:
the closer the particle is to the pore, the larger the
attractive force. 
(iii) Pore docking: 
The particle finally reaches
the pore entrance, 
where pore-particle molecular interactions
become relevant and often dominant.

EOF acts on both funnelling and pore docking phases. 
For highly charged molecules, such as DNA,
the main driver of the funneling phase is electrophoresis.
However, when the analyte 
has a low net charge, the 
electrophoretic force becomes negligible.
Proteins are a relevant class of biomolecules 
for which this is the case.
Indeed, they are either positively or negatively charged 
and, in general, their charge depends on the solution
pH so electrophoresis does not constitute a general
systematic method for protein capture in nanopores.
Therefore, an EOF directed toward the pore
appears as a promising alternative to favor capture.
Several works reported EOF induced capture for
proteins and peptides~\cite{huang2017electro,asandei2016electroosmotic,huang2020electro,schmid2021nanopore,saharia2021modulation,firnkes2010electrically}.
In Asandei {\sl et al.}~\cite{asandei2016electroosmotic},
a positively charged peptide 
was captured in a biological pore ($\alpha$HL), 
against electrophoresis. The EOF was tuned
by altering the protonation state 
of residues in the pore lumen 
decreasing the pH of the solution 
to 2.8 (see Fig.~\ref{fig:biological} for MD simulation data on the 
effect of pH in EOF through $\alpha$HL). 
Similar evidence has been reported in~\cite{soskine2012engineered}
where a globular protein is captured
by ClyA pore.
A recent approach to exploit EOF 
for protein capture and trapping
is to dock a DNA-origami sphere onto a solid-state nanopore.
The sphere partially seals off the nanopore and,
altering the distribution of ions in the pore 
region, gives rise to an EOF~\cite{schmid2021nanopore}.
The employment of EOF to enhance the capture of the analyte
is not limited to proteins.
EOF has proven relevant for 
sensing relatively large analytes 
with low charge, such as viruses in solid-state 
pores~\cite{arima2018selective}, 
as well as antibiotics and sugars in biological 
pores~\cite{bafna2020electroosmosis,bhamidimarri2016role,boukhet2016probing}.

In experimental studies on
molecule capture in nanopores, the 
role of EOF (and electrophoresis) on funneling
and pore docking phases can hardly be 
disentangled. For biological pores, for instance,
any mutation on the pore entrance aimed at altering 
the interaction between
pore and molecule may also change the EOF. 
In this respect,
MD simulations may help 
to isolate the effect of EOF on
the molecule's entry,
as shown for instance in~\cite{prajapati2020voltage}
using free-energy calculations.

Analytic theoretical frameworks trying to 
model particle capture 
under the concurrent action of electric forces,
EOF advection, and Brownian motion have been 
proposed with the aim of understanding the 
overall dependence of the capture frequency 
as a function of the magnitude of the 
applied voltage~\cite{chinappi2020analytical,grosberg2010dna,muthukumar2010theory}.
These analytical approaches, based in essence on
the solution of the advection-diffusion equation,
can also be used to analyze the cooperation/competition between
electrophoresis and electroosmosis in different conditions,
Fig.~\ref{fig:tec}e.
Although these theoretical routes 
may provide useful guidelines
for preliminary designs of systems
enhancing the capture of the desired molecules,
a precise determination of the role 
of EOF in nanopore sensing 
would largely benefit from the  
employment of the various computational approaches briefly discussed
in the present review.

As a final remark, EOF may be employed also 
for translocation control.
Tsutsui {\sl et al.}~\cite{tsutsui2021field}
showed 
that it is possible to slow down the 
particle's translocation through a solid-state 
nanopore controlling the EOF via a voltage-gated mechanism.
Zhang {\sl et al.}~\cite{zhang2020electroosmotic}
showed that for a solid
state nanopore, the 
translocation of proteins
slows down when the pore diameter is larger than the pore length.
They attributed this effect to 
a reduction of the EOF in the center of the channel,
an effect that is evident in continuum 
simulations, see {\sl e.g.} Fig.~\ref{fig:cont}c.

\subsection{Role of EOF in blue energy and desalination}
\label{sec:blue}

Besides sensing, nanopores are key elements
of emerging technologies with promising outcomes to solve global challenges.
For example nanopore based-membranes 
can harvest blue energy,
a sustainable energy~\cite{siria2017new}
and 
can improve access to clean water through high-performance filtration and desalination~\cite{werber2016materials}. 

In blue energy harvesting, 
where a salinity gradient is converted into electric power, 
a membrane selective for 
positive (or negative) ions separates two reservoirs containing 
solutions with different salinities, {\sl e.g.} sea and fresh water~\cite{siria2017new,yip2016salinity}.
Ions diffuse from the high salinity to the 
low salinity chamber.
Assuming as in Fig.~\ref{fig:tec}b that 
the membrane is cation selective,
the positive charge flux will 
be larger than the negative one, 
resulting in a net electric current.
For desalination applications, instead, the membrane 
hinders the passage of all the 
ions~\cite{cohen2012water,heiranian2015water,werber2016materials}.
In such devices, a pressure is applied on the high concentration
chamber and fresh water is obtained on the 
other side of the membrane, see Fig.~\ref{fig:tec}c.  
A similar working principle is used to purify water from
contaminants in ultrafiltration 
processes~\cite{tu2020rapid}.
In both blue energy and 
water treatment systems, dispersed nanoparticles
can clog the pore (membrane fouling~\cite{yip2016salinity,she2016membrane}),
dramatically reducing the performance of real-life systems with respect to
laboratory set-ups. 
In this scenario, EOF may be used to
induce a flow to clean the membrane from particles 
(backwashing~\cite{bowen1994electroosmotic,jagannadh1996electrokinetics}).
The decrease of the membrane performance is also due to ion concentration polarization near 
the membrane that, again, 
can potentially be attenuated with EOF~\cite{chan2020reduced}.
It is worth noting that in these two technologies the voltage $\Delta V$ 
is not the only applied load to the systems and
also concentration and pressure differences between the 
two reservoirs
need to be considered, making the transport 
phenomenon more intricate~\cite{marbach2019osmosis}.

One central and open question for all these technological applications is:
Given an electrolyte solution containing a mixture of nanoparticles, how can we design 
the nanopore and tune the
operating conditions to facilitate the capture of 
specific nanoparticles and/or hinder undesired ones?
In answering this question, EOF plays a crucial role
since it acts on all the dispersed molecules regardless of their charge
and dipole, constituting a more generic way of manipulating
particles in nanopore systems.

\section{Future outlook}

In this review, we discussed mechanisms
to control EOF in nanopores,
the main computational techniques used to predict EOF
and some of its recent applications in sensing, energy and filtration.
While such applications are very promising, the understanding of the experimentally observed phenomena often requires computer simulations 
to catch the large variety of physical effects influencing EOF in nanopores.
Explaining experimental observations has been, up to now, the most common 
use of computer simulations of EOF when combined with experiments,
see, {\sl e.g.}~\cite{huang2020electro,qiu2018abnormal,laohakunakorn2015reversal,asandei2016electroosmotic}.
Yet quantitative prediction is crucial to address the challenge of designing new generation 
nanofluidic devices exploiting EOF.
From the computational point of view, this requires a careful 
selection of the model and the numerical technique 
and to take into account the relevant effects 
in the system for all parameters considered, 
while using a minimal amount of computational resources.
Atomistic simulations 
are already well established and, thanks to 
the spread of GPU computing, they will 
probably be more and more used in the future.
Indeed, even lab scale GPU workstations are now able to perform simulations 
involving hundreds of thousands of atoms.
This is usually sufficient 
to obtain a reliable prediction of EOF 
in relatively small pores such as $\alpha$HL or FraC.
Supercomputers are still needed for larger systems
and, in particular, for simulation campaigns 
aimed at exploring different parameters, such as
the effect of pore modifications or of electrolyte 
composition on EOF. 
Nowadays, typically these simulation campaigns require weeks 
or months and are mainly devoted to fundamental
research projects. They are not yet easily scalable to industrial developments
where fast 
quantitative prediction are needed to optimize device performance.
Hence, a real breakthrough in nanopore technology will only be achieved 
via computational approaches which enable to efficiently 
explore a huge variety of 
nanopore systems and operational conditions on standard 
workstations within a computational time of a few days.
Such tools will pave the way for modern design practices at the nanoscale,
harvesting modeling and 
optimization techniques currently only used at larger scales.  
In this respect, continuum and mesoscale modeling 
appear as the natural choice for fast simulation
strategies.  
Although no technique is nowadays able to
robustly guarantee an {\sl a priori} quantitatively accurate 
prediction for EOF in real nanopores,
some of the recent progress
 in including
nanoscale effects 
in continuum and mesoscale approaches 
discussed in this review 
are quite promising. We expect that
they will contribute to the development
of a new generation of computational tools for
nanopore design.